\documentclass[fleqn,usenatbib]{mnras}

\usepackage{newtxtext,newtxmath}

\usepackage[T1]{fontenc}

\DeclareRobustCommand{\VAN}[3]{#2}
\let\VANthebibliography\thebibliography
\def\thebibliography{\DeclareRobustCommand{\VAN}[3]{##3}\VANthebibliography}

\usepackage{graphicx}	
\usepackage{amsmath}	
\usepackage[version=4]{mhchem}
\usepackage{gensymb}
\defcitealias{venot_2012}{V12}
\defcitealias{venot_2019}{V19}

\title[Chemical Network Benchmark]{Benchmarking Two Chemical Networks used in General Circulation Models of Hot Jupiters}

\author[D. A. Christie et al.]{D. A. Christie,$^{1,2}$\thanks{E-mail: christie@mpia.de}
M. Zamyatina,$^{2}$
E. H\'ebrard,$^{2}$
T. M. Evans-Soma,$^{3,1}$
N. J. Mayne,$^2$
E. K. H. Lee,$^{4}$\newauthor
S.-M. Tsai,$^5$
D. E. Sergeev,$^6$
R. Veillet,$^2$
and K. Kohary$^{2}$ 
\\
$^{1}$Max Planck Institute for Astronomy, K\"onigstuhl 17, D-69117 Heidelberg, Germany\\
$^{2}$Department of Physics and Astronomy, Faculty of Environment, Science and Economy, University of Exeter, Exeter EX4 4QL, UK \\
$^{3}$School of Science, University of Newcastle, Callaghan NSW, 2308, Australia\\
$^{4}$Center for Space and Habitability, University of Bern, Gesellschaftsstrasse 6, CH-3012 Bern, Switzerland \\
$^{5}$Institute of Astronomy \& Astrophysics, Academia Sinica, Taipei 10617, Taiwan\\
$^{6}$School of Physics, University of Bristol, HH Wills Physics Laboratory, Tyndall Avenue, Bristol BS8 1TL, UK
}

\date{Accepted 2026 April 15. Received 2026 April 15; in original form 2025 November 29}

\pubyear{\the\year{}}

\begin{document}
\label{firstpage}
\pagerange{\pageref{firstpage}--\pageref{lastpage}}
\maketitle

\begin{abstract}
Chemical kinetics is becoming an increasingly vital component of hot Jupiter general circulation models (GCMs). Here we simulate the hot Jupiter WASP-96b using two chemical networks, a reduced chemical network frequently used in the GCM literature (which we refer to as V19) and a more recent effective network making use of tables of {\em net} reactions ({\sc minichem}), coupled to the same GCM in order to provide a robust benchmark. We find a numerical escape criterion used by the Unified Model chemical kinetics solver to stop integration for the duration of the chemical timestep, independent of the chemical network, results in artificial quenching, overestimating of \ce{HCN}, \ce{CH4}, and \ce{NH3} abundances by factors of 1.5 to 3. With this criterion disabled, agreement between the two networks is improved, except for \ce{HCN} and \ce{NH3}, where different reaction rates and included species results in lower abundances in the V19 network.  While many rates differ between the networks, the lower quenched \ce{NH3} abundances in the V19 simulations are, in particular, due to the choice of \ce{NH2 + NH3 -> N2H3 + H2} reaction rate, which is poorly constrained in the literature.  This reaction also impacts the quenching of \ce{HCN}, which is additionally affected by the lack of \ce{CH2NH2} in the V19 network.   While there are reasons to favour the {\sc minichem} \ce{HCN} and \ce{NH3} abundances, ultimately, improved experimental and theoretical determination of reaction rates are needed to address the uncertainties and better characterize the quenching behaviour.
\end{abstract}




\section{Introduction}

The growth in our understanding of the atmospheres of hot Jupiters has necessitated the consideration that chemical abundances may not be in thermochemical equilibrium when modelling the atmospheres of these planets. The computational cost of solving networks consisting of hundreds of chemical species and thousands of reactions in a three-dimensional general circulation model (GCM) often requires simplifications to the network to make the problem tractable. The first attempt to address this in models of hot Jupiters was the relaxation method introduced in \citet{cooper_2006} which allowed for perturbations in \ce{CO} and \ce{CH4} abundances from chemical equilibrium to be investigated without the necessity of solving large chemical networks. This approach was expanded upon by \citet{tsai_2018} and \citet{mendonca_2018} and is still used in models of hydrogen dissociation in ultra-hot Jupiters \citep[see, e.g.,][]{tan_2019}. Another approach is to forego the costly solving of the chemical kinetics equations and instead model the assumed end result of the equations of chemical kinetics being integrated to a steady state. This approach is taken in \citet{steinrueck_2019} wherein they assume that abundances of \ce{CH4} and \ce{CO} are homogenized as a result of transport-induced quenching and perform a suite of simulations investigating different ratios of \ce{CH4} to \ce{CO}. More recent studies such as \citet{drummond_2020} and \citet{zamyatina_2023,zamyatina_2024} have taken advantage of a reduced network wherein only the most relevant reactions are included to reproduce abundances for species of interest \citep[see][for details of the methodology used in those works]{venot_2019}. This approach still involves integrations for a non-negligible number of species and reactions. The \citet[][hereafter V19]{venot_2019} reduced C/H/O/N network, for example, uses 30 species and 181 forward thermal reactions, which, while non-negligible in complexity, is significantly smaller than the recent updated network by \citet{veillet_2024} which has 175 species and 2550 reactions. The \citetalias{venot_2019} reduced network also omits photochemical reactions, which both reduces the size of the network and avoids the need to couple the radiative transfer scheme to the photolysis reactions.\footnote{ While general circulation models of hot Jupiters generally omit photochemistry, this has been investigated in the context of terrestrial exoplanets with oxidized atmospheres by \citet{ridgway_2023} and \citet{braam_2022}. } A method to further reduce the computational expense is to introduce {\em effective} or {\em net} reaction rates between species of interest, removing intermediate products in the network. This approach is taken in the thermochemical kinetics code {\sc minichem} \citep{tsai_2022}, which contains 12 species and 4 elementary forward reactions and 6 net forward reactions. This significantly smaller network makes {\sc minichem} attractive for chemical kinetics modelling in GCM simulations, which already suffer from significant computational overhead.  

To evaluate how {\sc minichem} may perform as a substitute for larger chemical networks, it is important to understand how it performs in the presence of forcings that push the gas away from chemical equilibrium. {\sc minichem} was initially benchmarked in one-dimensional models including vertical mixing parametrised by $K_{zz}$ against similar models involving the full {\sc vulcan} chemical network on which it is based and was found to perform well \citep{tsai_2022}. It has also been coupled to the {\sc exofms} GCM \citep{lee_2021} where it was used to simulate the atmospheres of WASP-39b and HD 189733b \citep{lee_2023}. In the case of HD 189733b, \citet{lee_2023} compare their results qualitatively to the results in \citet{drummond_2020} and \citet{zamyatina_2023}, both of which couple the \citetalias{venot_2019} chemical network to the {\sc unified model} GCM, and find good agreement. Direct comparison is difficult, however, as the GCMs solve different hydrodynamic equations and use different radiative transfer methods, initial conditions, planetary parameters, radiative coupling, and numerical damping schemes.

In this study, we ascertain the performance of {\sc minichem} versus the larger, reduced network of \citetalias{venot_2019} by coupling both to the {\sc unified model} GCM. This allows for the issues created by the differences listed above to be controlled for and the results more directly compared. It also provides an opportunity to benchmark the behaviour of the \citetalias{venot_2019} network and the chemistry solver implemented in the UM. 

The paper is structured as follows: in Section \ref{sec:methods}, the {\sc unified model} GCM is introduced, as are the chemical solvers to be investigated. The target planet -- WASP-96~b -- is also discussed. In Section \ref{sec:results}, the results of the suite of simulations covering a number of metallicities are compared and the performance of the two networks is quantified. The conclusions are summarised in Section \ref{sec:conclusions}.   The paper contains a number of appendices investigating the \ce{CH4} thermochemical data (Appendix \ref{Appendix:Methane}), the choices in reaction rates (Appendix \ref{Appendix:Rates}), chemical timestep (Appendix \ref{Appendix:Timestep}), and opacity dependence (Appendix \ref{Appdendix:Opacity}), as well as additional plots of interest (Appendix \ref{Appendix:Additional}). 


\section{Methods}
\label{sec:methods}
To create a self-consistent framework to compare the chemical kinetics networks and solvers, we couple both schemes to the Met Office's {\sc unified model} (UM) general circulation model. This approach controls for differences in dynamical cores within the GCMs as well as treatments in the radiative transfer solver.

\subsection{The Unified Model}
The UM solves the full, deep-atmosphere Euler equations \citep[see, e.g.,][]{mayne_2014a,wood_2014} and has been used extensively to model mini-Neptunes and hot Jupiters \citep[e.g.,][]{mayne_2019,drummond_2020,zamyatina_2023,christie_2021,christie_2022}. The radiative transfer is solved using the {\sc SOCRATES} radiative transfer code \citep{edwards_1996} which has been modified for use in hot Jupiter simulations by \citet{amundsen_2016,amundsen_2017} with the stellar irradiation modelled using a pseudo-spherical approximation \citep{jackson_2020}. The sources of opacity are \ce{H2O}, CO, \ce{CO2}, \ce{CH4}, \ce{NH3}, HCN, Li, Na, K, Rb, and Cs as well as collision-induced absorption by \ce{H2}-\ce{H2}\, and \ce{H2}-He and Rayleigh scattering by \ce{H2} and \ce{He}. The correlated-k opacity data for each species are combined during the simulation using equivalent extinction, allowing for the impact of the chemistry to feed back via changes in opacity. The volume mixing ratios of the non-alkali species are taken from the chemical network. As the alkali species Li, Na, K, Rb, and Cs are not in either chemical network investigated here, we assume them to be in chemical equilibrium and their abundances are computed using a look-up table generated using the equilibrium solver. This differs from previous investigations using the UM where the abundances of atomic alkali species are instead estimated using the threshold method outlined in \citet{amundsen_2016}, and this approach allows for metallicity to be appropriately considered in determining atomic alkali abundances.  The chemical mass fractions are stored as tracers, one for each species, with the dynamical core of the UM handling the advection of these tracers.  No diffusion of the tracers is applied, except for the implicit diffusion in the transport scheme. 

\subsection{Chemistry}
\label{Sec:Chem}
The goal of this work is to compare two approaches to thermochemical kinetics in the UM. Each simulation is initialized with the chemical abundances in equilibrium using the {\sc UM}'s Gibbs minimization solver \citep{drummond_2018} with the elemental abundances taken from \citet{caffau_2011} (C, N, O, and K) and \citet{asplund_2009} (Na, Li, Cs, Rb, and He). The chemical solvers are called every 125 dynamical timesteps ($\Delta t_\mathrm{chem}=3750\, \mathrm{s}$). This is consistent with the simulations of \citet{drummond_2020} and \citet{zamyatina_2023} but is less frequent than in \citet{lee_2023} where a chemical timestep of $1500\,\mathrm{s}$ is used. For a more direct comparison, we opt to use a chemical timestep of $3750\,\mathrm{s}$ in both cases.  We investigate the impact of this assumption in Appendix \ref{Appendix:Timestep} and find that differences resulting from shorter chemical timesteps are localized around sharp gradients in chemical abundances and are unlikely to impact the dynamics or synthetic observables. 

As discussed above, the chemical abundances inform the gas opacity, and thus can result in changes in gas temperature between different chemical schemes.  We opt to compare chemistry schemes with this effect included, as opposed to having the opacities be independent of the local chemical abundances as determined by the chemical kinetics solver, in order to understand differences in the chemistry schemes as they are used in the UM.  As noted below in Section \ref{Sec:thermal}, the temperature differences between the chemical kinetics schemes are small, and thus we do not expect temperature differences to significantly reinforce or exacerbate chemical abundance differences.  We investigate this possibility in Appendix \ref{Appdendix:Opacity}.

\subsubsection{The {\sc um} Chemical Kinetics Solver}
\label{sec:umvenot}
The chemical kinetics solver in the UM was developed in \citet{drummond_2020} and is based on the chemical kinetics solver used in the radiative-convective equilibrium code {\sc atmo} \citep{tremblin_2015,drummond_2016}. The forward reactions are taken from the reduced network of \citetalias{venot_2019} with reverse reactions computed using a mix of fits to JANAF \citep{chase_1998} and NASA Glenn \citep{mcbride_1994,mcbride_2002} thermochemical data. The reduced \citetalias{venot_2019} chemical scheme was developed from the full \citet{venot_2012} chemical scheme in the aim to reproduce accurately the observed abundances of key chemical species (\ce{H2O}, \ce{CH4}, CO, \ce{CO2}, \ce{NH3}, and HCN). The numerical integration is done using the {\sc dlsodes} ordinary differential equation (ODE) solver \citep{hindmarsh_2019,drummond_2020}. While the solver can run with different chemical networks, all UM exoplanet kinetics studies have used the \citetalias{venot_2019} chemical network, and \citetalias{venot_2019} is therefore the chemical network investigated here. As simulations done here will involve both the UM chemistry solver and the \citetalias{venot_2019} chemical network, and thus use assumptions made in each, we refer to this combination as ``UM/Venot'' moving forward.

The UM chemistry solver, as used previously, employs an {\em escape condition} in the integration of the equations of chemical kinetics. This is inherited from the {\sc atmo} code where it is used to determine convergence\footnote{{\sc vulcan} uses a similar convergence condition, as outlined in \citet{tsai_2017}.}; however, in the UM, it is used, in theory, to prevent unnecessary computational expense by halting the integration of the chemical kinetics in cells where the abundances are not expected to change appreciably over the chemical timestep, such as in the case where the abundances are in chemical equilibrium\footnote{We opt to use the term ``escape condition'' over ``termination condition'' to avoid confusion with scenarios in {\sc atmo} or {\sc vulcan} where meeting the criteria results in the simulation ending.}. The trigger for the escape condition, in practice, is implemented as follows. The chemistry is integrated forward in blocks of $\Delta t_\mathrm{chem}$, with the chemical kinetics routine taking smaller internal timesteps of $\delta t_\mathrm{chem}$ determined by {\sc dlsodes}. Taking the integration to start at $t=0$, after $t=1\,\mathrm{s}$ the chemical kinetics routine checks the rate of change. For a reference time $t_0$, initially set to $1\,\mathrm{s}$, the chemical kinetics routine computes for each species the quantity

\begin{equation}
\Delta_s = \frac{n_s(t) - n_s(t_0)}{n_s(t_0)}\,\, .
\end{equation}

\noindent If, on three consecutive internal timesteps $\delta t_\mathrm{chem}$, the code satisfies both $\Delta_s \leq 10^{-2}$ and $\Delta_s/(t-t_0) \leq 10^{-4}$ for all species $s$, the routine assumes the abundances will not evolve substantially over the chemical timestep and the integration of the chemical kinetics terminates until the next chemical timestep. The reference time $t_0$ is updated every 10\% increase in integration time of the chemical kinetics solver, (i.e., when $t \geq 1.1 t_0$, $t_0 \rightarrow 1.1 t_0$). As this escape condition artificially retards the evolution of the chemical abundances, it has the potential to increase quenching within the atmosphere, as choosing to not integrate forward the abundances is equivalent to assuming an infinitely long chemical timescale and thus significantly longer than any finite dynamical timescale. When this escape condition is not used, the chemical kinetics are simply integrated forward for the full chemical timestep $\Delta t_\mathrm{chem}$ without any possibility of early termination. We test both simulations with and without the use of this escape condition in the analysis below.  We do not, however, test stricter tolerances on the escape condition which may have the effect to reduce artificial quenching that may result.

\subsubsection{{\sc minichem}}

The {\sc minichem}\footnote{All simulations involving {\sc minichem} were done with commit e354183 from the repository available at \href{https://github.com/ELeeAstro/mini_chem}{https://github.com/ELeeAstro/mini\_chem}. While minor changes were made to allow the code to be coupled to the UM, no changes were made to the solver itself. } solver \citep{tsai_2022,lee_2023} solves for the evolution of H, \ce{H2}, \ce{He}, \ce{H2O}, O, OH, CO, \ce{CO2}, \ce{CH4}, \ce{C2H2}, \ce{NH3}, and HCN. With the exception of \ce{C2H2}, these species are all in the \citetalias{venot_2019} network. The network consists of four elementary forward reactions,

\begin{align}
\mathrm{OH} + \mathrm{H_2} &\rightarrow \mathrm{H_2O} + \mathrm{H}\nonumber\\
\mathrm{OH} + \mathrm{CO} &\rightarrow \mathrm{H} + \mathrm{CO_2}\nonumber\\
\mathrm{O} + \mathrm{H_2} &\rightarrow \mathrm{OH} + \mathrm{H}\nonumber\\
\mathrm{H} + \mathrm{H} + \mathrm{M} &\rightarrow \mathrm{M} + \mathrm{H_2}\,\, ,\nonumber
\end{align}

\noindent supplemented with six net forward reactions,

\begin{align}
\mathrm{CH_4} + \mathrm{H_2O} &\rightarrow \mathrm{CO} + \mathrm{3H_2}\nonumber \\
\mathrm{2CH_4} &\rightarrow \mathrm{C_2H_2} + \mathrm{3H_2}\nonumber \\
\mathrm{CO} + \mathrm{CH_4} &\rightarrow \mathrm{C_2H_2} + \mathrm{H_2O}\nonumber \\
\mathrm{2NH_3} &\rightarrow \mathrm{N_2} + \mathrm{3H_2}\nonumber \\
\mathrm{CH_4} + \mathrm{NH_3} &\rightarrow \mathrm{HCN} + \mathrm{3H_2}\nonumber \\
\mathrm{CO} + \mathrm{NH_3} &\rightarrow \mathrm{HCN} + \mathrm{H_2O}\,\, .\nonumber 
\end{align}

\noindent The net forward reaction rates are available in a look-up tables depending on pressure and temperature, with lookup tables available for different metallicities. The reverse reaction rates are then computed using the supplied thermodynamic data. A more in-depth discussion of the computation of the net rates and the limiting reactions can be found in \citet{tsai_2022}.

The {\sc minichem} code provides a number of ODE solvers that can be used. For the purposes of this investigation, we use {\sc seulex} \citep{hairer_2010}, following \citet{lee_2023}. When this project was initiated, {\sc dlsodes} support was not included in {\sc minichem}, thus the choice was made to use the solver used in a previous coupling to a GCM. We note that {\sc minichem} now offers {\sc dlsodes} support.

\subsubsection{Thermochemical Data}

While forward reaction rates are explicitly provided, reverse reaction rates are computed using thermochemical data, and these data are sometimes left implied or ambiguous. {\sc minichem} uses 9-term NASA Glenn fits \citep[][see also the {\sc minichem} repository]{mcbride_2002}, while the \citetalias{venot_2019} and \citet{venot_2012} networks use a mix of fits to the JANAF \citep{chase_1998} data, fits from \citet{burcat_2005}, and 7- and 9-term NASA Glenn fits \citep{mcbride_1994,mcbride_2002}, with the UM implementation using these same data.  An exception to this is thermochemical data for \ce{N2H3}, which is not provided in any of the previously mentioned sources.  The provenance is unknown, but the fit used generally agrees with the fits of \citet{burcat_2005}, although the coefficients are not identical.

All of these data generally agree between \citetalias{venot_2019} and {\sc minichem}, with a notable exception being the \ce{CH4} data, where the fits diverge above 1000~K. The Venot networks use 7-term fits to enthalpies computed assuming the rigid-rotator harmonic-oscillator (RRHO) approximation, consistent with \citet{burcat_2005}, while the 9-term NASA Glenn fits \citep{mcbride_2002} used by {\sc minichem} are fit to enthalpies computed using the non-rigid-rotator anharmonic-oscillator (NRRAO) approximation (see also \citealt{burcat_2005}). This will result in differences in the equilibrium abundances of \ce{CH4} in the deep atmosphere where the temperatures exceed 1000~K, on the order of a few percent, and the impact on the simulations here is discussed in Section \ref{sec:methane} while a more general discussion of the impact on equilibrium abundance of \ce{CH4} is found in Appendix \ref{Appendix:Methane}. In an analysis of available polynomial fits to thermochemical data as well as their own fits, \citet{wang_2023} recommend the \citet{mcbride_2002} fits for \ce{CH4}, and thus the deep atmosphere \ce{CH4} abundances in the UM/Venot simulations are, in the context of this chemical network comparison, an likely an underestimation. 

For the other species shared by both \citetalias{venot_2019} and {\sc minichem} and are considered by \citet{wang_2023}, they generally recommend the NASA Glenn fits, with the exceptions being \ce{N2}, \ce{CO2}, \ce{HCN}, and \ce{H2O}.  For \ce{N2}, they recommend fits to the JANAF data, although these differ by less than 1\% from the NASA Glenn fits over the prescribed temperature range.  For \ce{CO2}, the JANAF fit is similarly recommended, with the fit differing by less than 1\% below 4000~K and diverging from the NASA Glenn fit only above this temperature. In this case, however, both \citetalias{venot_2019} and {\sc minichem} use the NASA Glenn fits, and thus the difference will not impact the comparison.  For both \ce{HCN} and \ce{H2O}, \citet{wang_2023} recommend their own fits generated from partition functions constructed from ExoMol data. For \ce{HCN}, the JANAF and NASA Glenn fits differ by less than 0.15\%, and thus should not influence the comparison between the two networks here; however, we note that the ExoMol fit in \citet{wang_2023} differs significantly beginning around 1000~K, with an inversion in the specific heat capacity around 2000~K.   We note this for future modelling efforts, but as neither network considered here uses this fit, we do not investigate it further.  Lastly, the \ce{H2O}, the fits to the JANAF tables, used by \citetalias{venot_2019}, and NASA Glenn fits differ by less than 1\% below 2000~K, and less than 2\% below 4000~K. The recommended ExoMol fit is bounded by these two fits, and thus we expect the differences to be relatively minor, and as neither network here uses the ExoMol fit, it does not impact the comparison.

\subsection{The Target: WASP-96b}
To understand the behaviour of these chemical networks and their reproduction of chemical quenching due to global transport processes in hot Jupiter atmospheres, we test the chemical kinetics schemes in simulations of WASP-96b \citep{hellier_2014} as it has previously been the subject of an investigation of chemical kinetics in its atmosphere using the UM \citep{taylor_2023,zamyatina_2024} and was also a target of JWST Early Release Observations \citep[ERO;][]{radica_2023}. While observations indicate the planet is roughly $10\times$ solar metallicity, we simulate three different cases: $0.1\times$, $1\times$, and $10\times$ solar, based on the available {\sc minichem} rate tables that may be used for hot Jupiters.\footnote{{\sc minichem} offers tables for $100\times$ and $500\times$ solar metallicity as well, although we do not test these higher metallicities.}

\subsubsection{Generating the Initial Temperature Profile}
Simulations were initialised with a hydrostatic atmosphere using 1D temperature profiles and no winds.  As the equatorial jets in close-in, gaseous exoplanets cause heating of the deep atmosphere \citep{amundsen_2016,tremblin_2017,sainsbury-martinez_2019}, 1D radiative-convective equilibria do not represent equilibrium states in 3D models of these planets. In simulations that are initialized using temperature-pressure profiles in 1D radiative-convective equilibrium, known as a {\em cold start} initial conditions, this heating in the deep atmosphere results in a temperature inversions as the integration times required to heat the deep atmosphere to an equilibrium state are often prohibitive.  To avoid these issues associated with cold start simulations, the initial pressure-temperature profiles are generated two stages. We first run a one-dimensional {\sc atmo} simulation for the desired metallicity using equilibrium chemistry and including the species from the \citetalias{venot_2019} network plus the alkali species in the lookup tables. This temperature profile, shifted by $600\,\mathrm{K}$ following \citet{amundsen_2016}, is then used as the initial {\em hot start} profile for a UM equilibrium chemistry simulation which is run for $1000$ Earth days. The global average temperature profile at the end of the simulation is then the initial PT profile that we use going forward for all simulations with the prescribed metallicity.  This approach will better capture the temperature in the deep atmosphere, avoiding the deep atmosphere temperature inversions seen in \citet{zamyatina_2024}; however, the extremely long convergence times in the deep atmosphere mean it is still unlikely to be converged \citep{sainsbury-martinez_2019}, and continued evolution of the deep atmosphere temperature structure over the course of the simulations presented below, while not as dramatic as in {\em cold start} simulations, still occurs.

The continued evolution of the deep atmosphere will also result in a discrepancy in the excess flux from the top of the atmosphere. All simulations here assume an intrinsic temperature $T_\mathrm{int}$ of $100\, \mathrm{K}$ corresponding to an excess flux $\sigma T_\mathrm{int}^4$, with the flux entering across the inner computational boundary from the unmodelled interior.  As the atmosphere is not in thermal equilibrium and continues to evolve, this excess at the inner boundary will not necessarily be identical to the excess flux at the outer boundary.


\subsubsection{The Suite of Simulations}

The simulation parameters are presented in Table \ref{Tbl:Common}. As the change in metallicity alters the mean molecular mass and thus the scale height and as the different metallicities have different initial temperature profiles, the height of the computational domain and the inner boundary radius are adjusted for each metallicity to roughly match the observed transit radius.   All simulations, both equilibrium and chemical kinetics, are initialised using the temperature profiles discussed in the previous section, with each simulation run for 1000 days, not including the simulation time used in generating the initial profile. 

As we are simulating multiple metallicities each with multiple chemistry schemes for the same target, the number of simulations and computational cost becomes large, making the investigation of other targets prohibitive.  This restrictions the parameter space investigated and thus, for example, the behaviour of the schemes for ultra-hot and temperate Jupiters is uncertain.

\begin{table*}
\caption{Simulation Parameters}
\label{Tbl:Common}
\begin{tabular}{lcccc}
\hline
\hline
 & $0.1\times$ Solar & $1\times$ Solar & $10\times$ Solar & Units\\
\hline
{\em Grid and Domain} \\
Longitude Cells & \multicolumn{3}{c}{144} & \\
Latitude Cells & \multicolumn{3}{c}{90} & \\
Vertical Layers & \multicolumn{3}{c}{86} & \\
Domain Height & $1.03\times 10^7$ & $1.03\times 10^7$ & $1.01\times 10^7$ & m \\
Domain Inner Radius & $7.936\times 10^7$ & $7.936\times 10^7$ & $7.86\times 10^7$ & m \\
Hydrodynamic Timestep & \multicolumn{3}{c}{30} & s \\
Simulation Duration & \multicolumn{3}{c}{1000} & Earth days \\
\\
{\em Radiative Transfer} \\
Wavelength Bins & \multicolumn{3}{c}{32} & \\
Wavelength Minimum & \multicolumn{3}{c}{0.2} & $\mathrm{\mu m}$\\
Wavelength Maximum & \multicolumn{3}{c}{322} & $\mathrm{\mu m}$\\
Radiative Timestep & \multicolumn{3}{c}{150} & s\\
\\
{\em Damping and Diffusion} $^\mathrm{a}$ \\
Damping Profile & \multicolumn{3}{c}{Horizontally Uniform} & \\
Damping Coefficient & \multicolumn{3}{c}{0.15} & \\
Damping Depth ($\eta_s$) & \multicolumn{3}{c}{0.8} & \\
Diffusion Parameter ($t_K$) & & 4 & \\
\\
{\em Planet}\\
Intrinsic Temperature & \multicolumn{3}{c}{100} & K \\
Initial Inner Boundary Pressure & \multicolumn{3}{c}{200} & bar\\
Semi-major axis $a$ & \multicolumn{3}{c}{$4.53\times 10^{-2}$} & AU \\
Stellar Constant at 1 AU & \multicolumn{3}{c}{1272} & $\mathrm{W\, m^{-2}}$ \\
Specific gas constant R & 3558.0 & 3516.3 & 3164.7 & $\mathrm{J\,kg^{-1}K^{-1}}$ \\
Specific heat capacity $c_\mathrm{P}$ & 12713.6 & 12625. & 11476.7 & $\mathrm{J\,kg^{-1}K^{-1}}$\\
g at inner boundary & 9.85 & 9.85 & 10.04 & $\mathrm{m\,s^{-2}}$\\
\hline
\multicolumn{5}{l}{$^\mathrm{a}$  {\footnotesize See \citet{christie_2024} for a discussion of damping and diffusion parameters.}}
\end{tabular}
\end{table*}

\section{Results and Discussion}
\label{sec:results}

For each of the three metallicities investigated -- $0.1\times$, $1\times$, and $10\times$ solar -- we ran simulations with {\sc minichem} as well as UM/Venot, and for the latter we tested simulations with and without using the escape condition in the solver. In all cases, of the opacity sources typically observed in hot Jupiter atmospheres, we find the \ce{H2O} and \ce{CO} remain near chemical equilibrium throughout the simulations, while \ce{CH4}, \ce{CO2}, \ce{HCN}, and \ce{NH3} all exhibit departures from equilibrium.

\subsection{Thermal Structure}
\label{Sec:thermal}
As the abundances of opacity sources depends on the chemical kinetics, it is possible for temperatures to differ between simulations with different chemistry schemes; however, in practice, the dominant contributors to the gas phase opacity -- \ce{H2O} and \ce{CO} -- do not vary significantly and do not show departures from equilibrium, so the temperature differences are ultimately small. We do, however, note them here.

\begin{figure*}
	\includegraphics[]{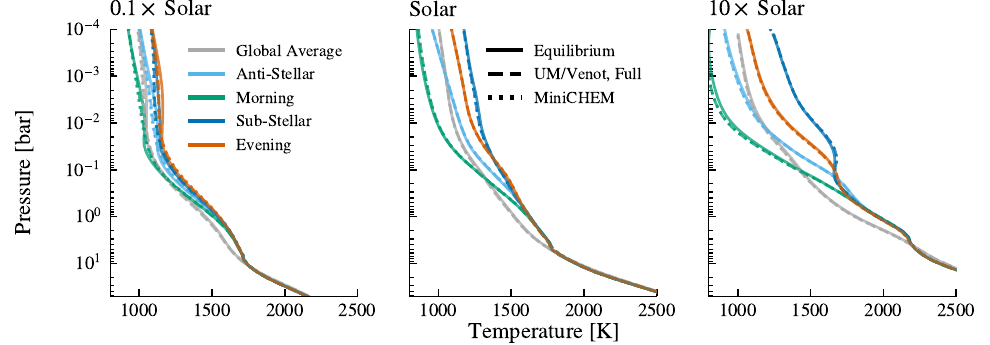}
    \caption{Equatorial pressure-temperature profiles at the substellar, anti-stellar, morning, and evening terminators, as well as the area-weighted global average profile for the equilibrium (solid lines), UM/Venot (dashed lines) and {\sc minichem} simulations (dotted lines).  }
    \label{Fig:pt}
\end{figure*}

Figure \ref{Fig:pt} shows the area-weighted pressure-temperature profiles for each of the simulations as well as the equatorial profiles at the substellar and anti-stellar points as well as along the morning and evening terminators. The area-weighted profile differs by at most 8~K, 4.5~K, and 0.2~K for the $0.1\times$, $1\times$, and $10\times$ solar metallicity cases, respectively, while the individual equatorial profiles differ by no more than 10~K, 6~K, and 11~K. While the simulations are only run for 1000~days and the profiles will likely continue to evolve, especially in the deep atmosphere \citep[see, e.g.,][]{sainsbury-martinez_2019}, the agreement in the temperature profiles for the different chemical schemes does add a measure of confidence that they will not diverge significantly over typical simulation timescales.

\begin{figure*}
	\includegraphics[]{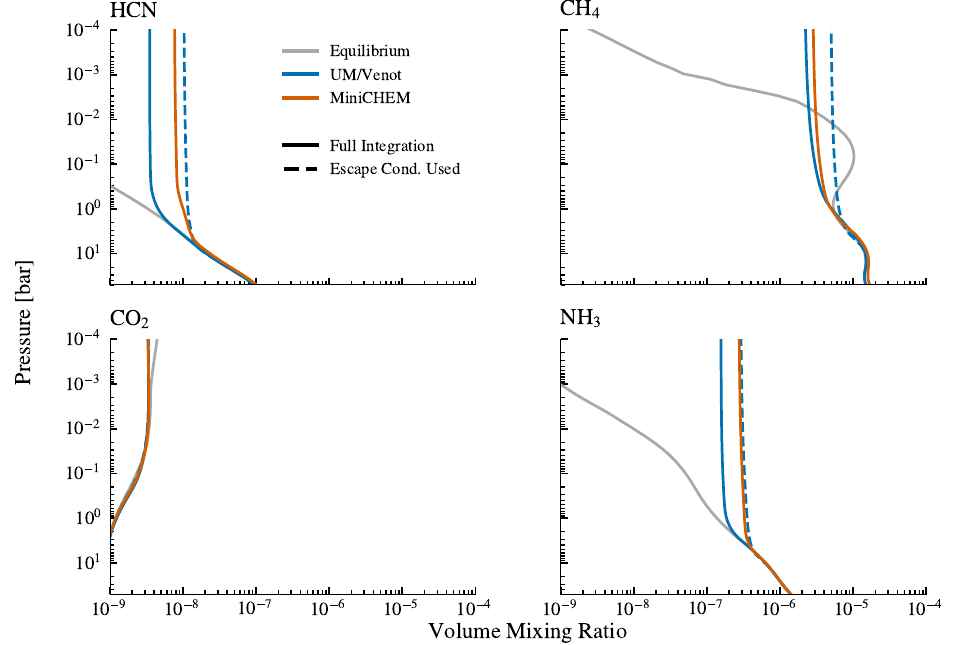}
    \caption{Area-weighted mass mixing ratios as a function of pressure for species that show noticeable amounts of quenching in the $0.1\times$ solar metallicity cases.}
    \label{Fig:chem_mdhm1}
\end{figure*}

\begin{figure*}
	\includegraphics[]{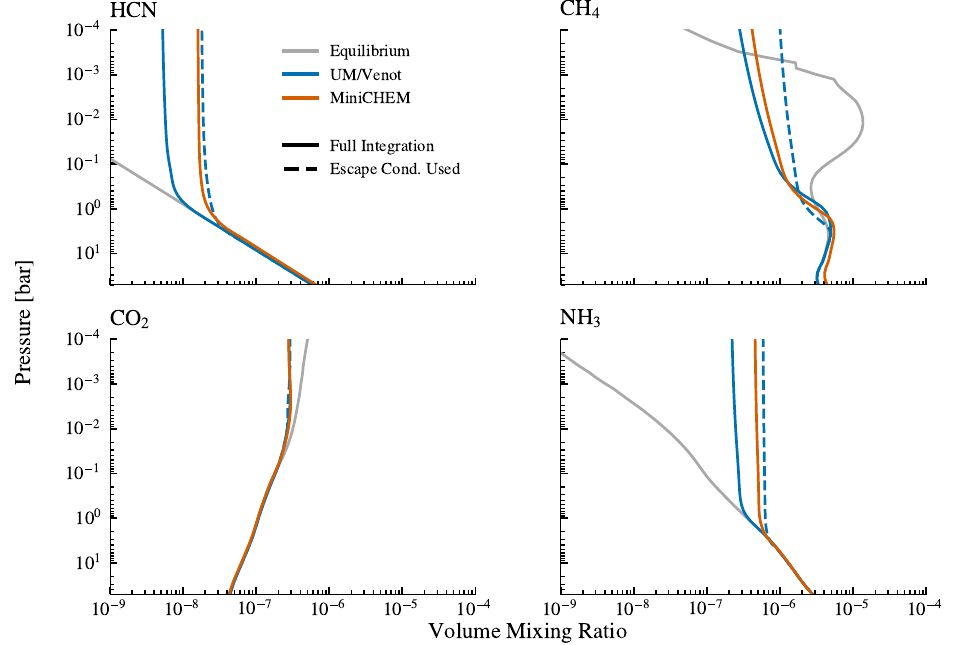}
    \caption{Area-weighted mass mixing ratios as a function of pressure for species that show noticeable amounts of quenching in the solar metallicity cases. }
    \label{Fig:chem_mdh0}
\end{figure*}

\begin{figure*}
	\includegraphics[]{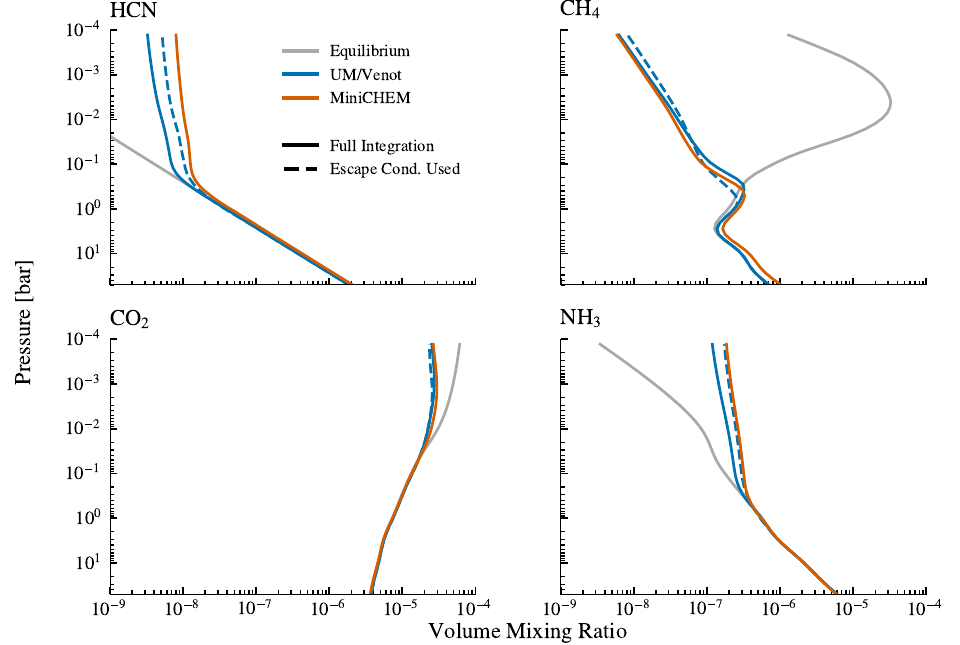}
    \caption{Area-weighted mass mixing ratios as a function of pressure for species that show noticeable amounts of quenching in the $10\times$ solar metallicity cases. }
    \label{Fig:chem_mdh1}
\end{figure*}

\subsection{The Impact of the Escape Condition}

We first compare UM/Venot simulations done with and without the escape condition before comparing to the {\sc minichem} results. The area-weighted chemical abundances for $0.1\times$, $1\times$, and $10\times$ solar metallicities are shown in Figures \ref{Fig:chem_mdhm1}, \ref{Fig:chem_mdh0}, and \ref{Fig:chem_mdh1}, respectively. For each of the metallicities tested, \ce{HCN}, \ce{CH4}, and \ce{NH3} quench deeper in the atmosphere when the escape condition is used, although the quench pressure differs by less than an order of magnitude. Due to the sharp gradient of the equilibrium abundance profiles, this results in an overestimation of the \ce{HCN}, \ce{CH4}, and \ce{NH3} volume mixing ratios by factors of $1.5$ to $3$, with the errors becoming smaller for larger metallicities. \ce{CO2} does not show the same behaviour, with the simulations using the escape condition only differing by 10\% from those using full integration for the $0.1\times$ and $1\times$ solar metallicities and differing by at most 40\% for the $10\times$ solar metallicity case. This increase in error in the \ce{CO2} abundance with metallicity may be due to the lower metallicity cases showing less quenching of \ce{CO2} in general, although this remains speculation.

For $1\times$ and $10\times$ solar metallicities, \ce{CH4} volume mixing ratios are overestimated at low pressures; however, around $\sim 1\,\mathrm{bar}$, the simulations using the escape condition exhibit lower abundances than the simulations which did not use it due to the equilibrium abundance profile of \ce{CH4} having an inversion at these pressures. The quenched abundance profiles are thus more complicated at these pressures than would be found for a monotonic equilibrium abundance profile, however, the result remains the same that the quenched abundances are overestimated when the escape condition is used.

\subsection{Comparing {\sc minichem} and UM/Venot}

We now compare the UM/Venot simulations without the escape condition with the {\sc minichem} simulations.

\begin{figure*}
	\includegraphics[]{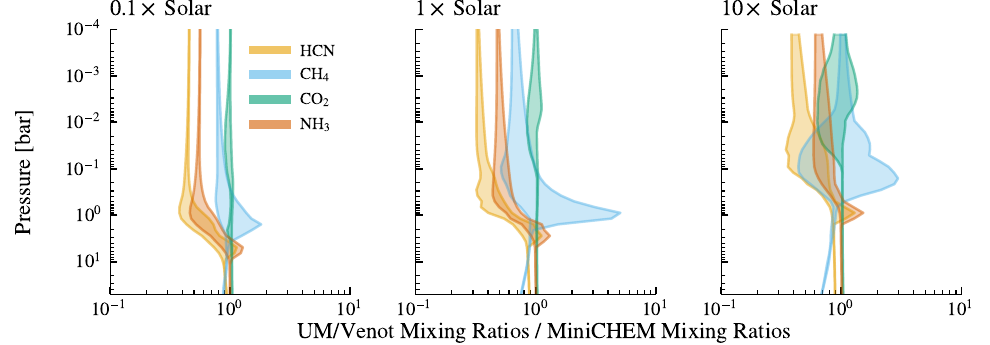}
    \caption{The range of values at the end of the simulations for the local ratio of the UM/Venot and {\sc minichem} volume mixing ratios (VMR) at given pressures.  The ratio of VMRs is evaluated at every point on the latitude-longitude grid to construct the range of values. Values greater than unity thus indicate local excesses in the UM/Venot VMR relative to the {\sc minichem} VMR while values less than unity indicate a local excess in the {\sc minichem} VMR relative to the UM/Venot VMR.}
    \label{Fig:chem_diff}
\end{figure*}

\subsubsection{Methane in the Deep Atmosphere}
\label{sec:methane}

Before comparing the quenching properties and divergences from chemical equilibrium, we first note differences in the deep atmosphere abundances of \ce{CH4}. While the UM/Venot simulations agree with the equilibrium simulations in the deep atmosphere, this is due to the use of the same thermochemical data in the equilibrium simulations and the UM/Venot chemical kinetic simulations. Both UM/Venot and {\sc minichem} chemical kinetic simulations are in equilibrium in the deep atmosphere with respect to their individual thermochemical data\footnote{Test equilibrium simulations using the {\sc minichem} thermochemical data were done to confirm that this is, in fact, the case.}. 

These deep atmospheric differences are small, less than a factor of 2 in abundance, as the thermochemical data for \ce{CH4} only begin to differ above $1000\,\mathrm{K}$. While it is beyond the scope of this paper, it has been proposed by \citet{fortney_2020} to use methane abundance at the photosphere to probe the conditions of the deep atmosphere, and this has been done in the cases of GJ~436b \citep{agundez_2014,morley_2017} and WASP-107b \citep{welbanks_2024,sing_2024}. Higher temperatures in the deep atmosphere deplete \ce{CH4} which, through vertical mixing and transport-induced quenching, can alter the photospheric abundance of methane. Accurate calculations of deep atmosphere \ce{CH4} abundance, and thus accurate thermnochemical data, are essential to constraining the temperatures in the deep atmosphere using this method.

\subsubsection{Main Opacity Sources}

For all three metallicities investigated here, the most abundant opacity sources, after \ce{H2} and \ce{He}, are \ce{CO} and \ce{H2O}. All remain relatively constant in abundance, with differences between the equilibrium simulation and either the UM/Venot or {\sc minichem} chemical kinetics simulation less than 17\%, 3.0\%, and 1.4\% for $0.1\times$, $1\times$, and $10\times$ solar metallicity, respectively, with the largest differences occurring at pressures less than 0.1~bar where the equilibrium abundances vary with pressure and transport-induced quenching homogenises the vertical profile. The UM/Venot and {\sc minichem} simulations differ from each other by less than 1.7\%, 0.25\%, and 1.2\% for $0.1\times$, $1\times$, and $10\times$ solar metallicity, respectively, thus any differences due to chemical network choice, at least for \ce{CO} and \ce{H2O}, are smaller than differences due to the departure from chemical equilibrium.

\subsubsection{Significantly Quenched Opacity Sources}

The opacity sources \ce{HCN}, \ce{CH4}, \ce{CO2}, and \ce{NH3} all depart from their equilibrium abundances due to transport-induced quenching (see solid blue and red lines in Figures \ref{Fig:chem_mdhm1} to \ref{Fig:chem_mdh1}). For \ce{CO2} and \ce{CH4}, the global mean profiles are within 60\% of each other, although we note that for \ce{CH4}, the local differences, shown in Figure \ref{Fig:chem_diff}, can be larger. The largest local differences occur around the quench pressures due to the UM/Venot and {\sc minichem} simulations departing from equilibrium at slightly different pressures. 

The larger differences observed in \ce{HCN} and \ce{NH3} quenching between UM/Venot and {\sc minichem}, observed for all metallicities, with differences in the global mean profiles between factors of 1.5 and 3, are likely due to the differences in nitrogen chemistry between the {\sc vulcan} chemical network that underlies {\sc minichem} and \citetalias{venot_2019}\footnote{We note that UM/Venot simulations using the escape condition tend to agree better with the {\sc minichem} results compared to those that don't.  This is a coincidence. }. Of the 83 forward reactions involving nitrogen in the \citetalias{venot_2019} network, only 16 are also in the {\sc vulcan} network and have the same rate.  {\sc vulcan} also includes reactions that are not included in \citetalias{venot_2019}. At pressures of 1 bar, \ce{CH2NH2 + H -> CH2NH + H2} is identified as the limiting reaction in the net reaction \ce{CH4 + NH3 -> HCN + 3H2} in the {\sc minichem} network (see the {\sc minichem} repository) which is absent from \citet{venot_2012,venot_2019}. This is supported by \citet{veillet_2024} who investigate differences in \ce{HCN} abundances within their network compared to \citet{venot_2020}, which updates \citet{venot_2012,venot_2019}, and conclude they are the result of the inclusion of \ce{CH2NH} and updated thermochemical data.

The differences in \ce{NH3} quenching between UM/Venot and {\sc minichem} are likely due to differences in the rates for the \ce{NH2 + NH2 -> N2H2 + H2} and \ce{NH2 + NH3 -> N2H3 + H2} reactions.  \citet{veillet_2024} find the updated rate for the former reaction impacts their results compared to the earlier \citet{venot_2020}. The rate is also updated in {\sc vulcan} to the rate in \citet{klippenstein_2009} and differs by orders of magnitude from the rate used in \citet{venot_2012,venot_2019}.\footnote{We note that both \citet{veillet_2024} and \citet{glarborg_2018} use the same rate from \citet{klippenstein_2009}, but with a typo in the exponent, potentially leading to differing results. The rate is correct in {\sc vulcan}.}  The latter reaction, \ce{NH2 + NH3 -> N2H3 + H2}, is not included in \citet{veillet_2024}, following its omission from the \citet{glarborg_2018} network, thus does not provide insights about any potential impact.  However, the forward rates for this reaction in \citetalias{venot_2019} and {\sc vulcan} differ by many orders of magnitude as well, and we find altering the rates of these two reactions is sufficient to alter \ce{NH3} and \ce{HCN} abundances and account for the difference in \ce{NH3} quenching between \citetalias{venot_2019} and {\sc minichem} (see Appendix \ref{Appendix:Rates}).   As discussed in \citet{moses_2014}, the appropriate rates for these reactions and reactions involving \ce{N2H_x} more generally remain uncertain, and a recent analysis by \citet{marshall_2023} estimated an upper limit on the rate orders of magnitudes smaller than the \citet{konnov_2000} rate, although this is in tension with contemporary analyses of experimental data by \citet{abian_2021} and \citet{manna_2023} (see the discussion in Appendix \ref{Appendix:Rates}). Adjudicating this is beyond the scope of this paper, and we instead take this as an uncertainty in the models.

\subsubsection{Radicals}

\begin{figure*}
	\includegraphics[]{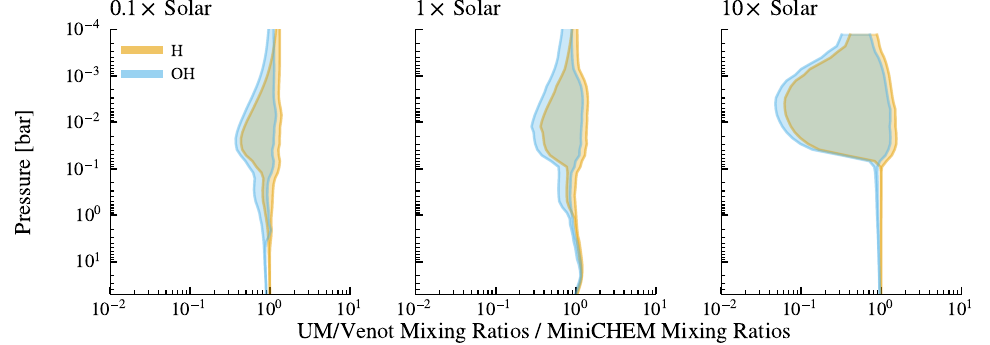}
    \caption{The same as Figure \ref{Fig:chem_diff}, except for the \ce{H} and \ce{OH} radicals.}
    \label{Fig:chem_diff_radicals}
\end{figure*}

\begin{figure*}
	\includegraphics[]{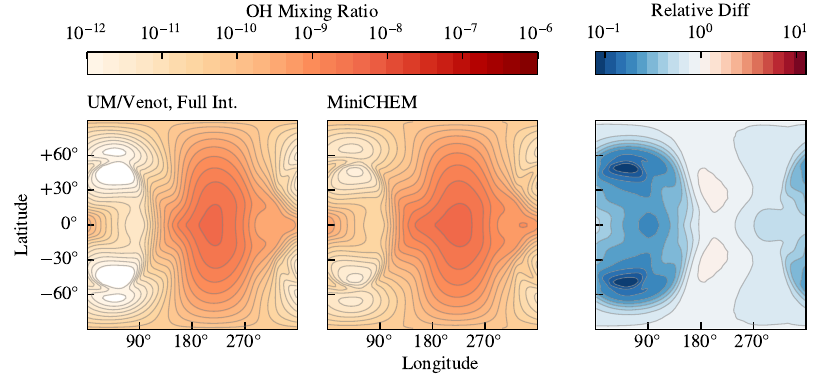}
    \caption{ {\em Left and middle panels: } The local \ce{OH} volume mixing ratio at 1 mbar for the $10\times$ solar metallicity UM/Venot simulation using full integration (left panel) and {\sc minichem} simulation (middle panel). {\em Right panel: } The relative difference between the two simulations, as computed as $\left(x_\mathrm{OH, UM/Venot} - x_\mathrm{OH, MiniCHEM}\right)/x_\mathrm{OH, UM/Venot}$. }
    \label{Fig:horiz_rad}
\end{figure*}

While the radicals in {\sc minichem}, \ce{H} and \ce{OH}, are not opacity sources in the GCM, they are instrumental in various reaction pathways, and were these networks to be coupled to photochemical networks, would be relevant in assuring accuracy. Both species exhibit quenching, and the local differences between the UM/Venot and the {\sc minichem} networks differ by up to an order of magnitude (Figure \ref{Fig:chem_diff_radicals}), with the largest differences between simulations being observed in the $10\times$ solar metallicity case. The largest differences occur on the nightside, as illustrated in Figure \ref{Fig:horiz_rad} for the $10\times$ solar metallicity cases.  This is caused by differing chemical timescales for the radicals as they are carried by winds from around the substellar point where they are near thermochemical equilibrium  to the nightside, with the advection timescale neither short enough to homogenise the abundances nor long enough for the radicals to return to equilibrium.  Two differences in the handling of radicals in the chemistry schemes combine to result in these discrepancies.  First the forward reaction rate for \ce{OH + H2 -> H2O + H} differs between \citetalias{venot_2019} and {\sc minichem} by roughly 10\% across the temperature ranges in the simulations.  Second, while \ce{H} and \ce{OH} are involved only in the four elementary forward reactions in {\sc minichem}, the \citetalias{venot_2019} network contains 49 forward reactions involving \ce{OH} and 67 forward reactions involving \ce{H}, with these additional reactions likely having a non-negligible impact.  As the \citetalias{venot_2019} network is more complete, we expect that for these trace species with fast reaction rates, it is thus more reliable.

For reference, we also include equatorial profiles along different directions in Figures \ref{Fig:chem_h} and \ref{Fig:chem_oh} of Appendix \ref{Appendix:Additional}.

\subsection{Impact on Transmission}

\begin{figure*}
	\includegraphics[]{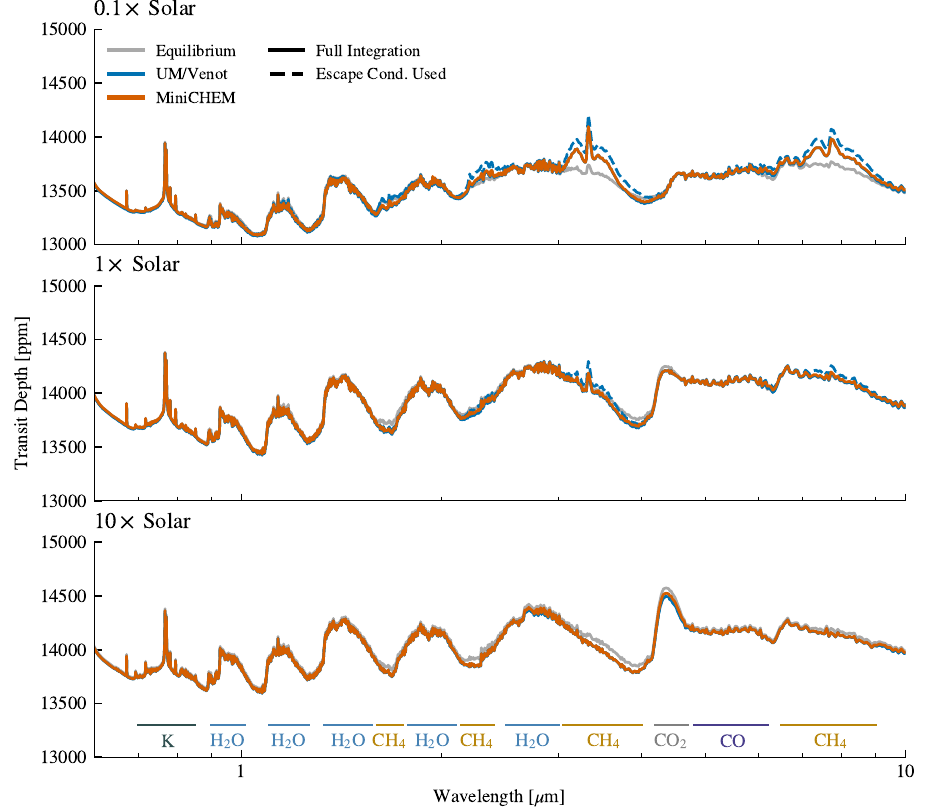}
    \caption{The transmission spectra for each of the metallicities investigated. In many cases, the UM/Venot simulations using the full integration overlap with the {\sc minichem} simulations to such an extent that the lines for the UM/Venot spectra cannot be discerned.  It is possible, however, to see the UM/Venot simulations using the escape condition due to the differing \ce{CH4} quenching behaviour.}
    \label{Fig:transmission}
\end{figure*}

While the focus here is on the chemical solvers themselves, we do investigate to what extent these differences impact the synthetic transmission spectra generated by the UM's post-processing routines \citep{lines_2018b}. The results are shown in Figure \ref{Fig:transmission} with the largest differences occurring in regions with strong \ce{CH4} features. Examining first the differences between the UM/Venot simulations with and without the escape condition, we find that the use of the escape condition results in differences of up to 107, 125, and 30 ppm for $0.1\times$, $1\times$, and $10\times$ solar metallicity, respectively, with the differences due to the increased quenching of \ce{CH4} in the simulations that use the escape condition (see Figures \ref{Fig:chem_mdhm1} to \ref{Fig:chem_mdh1}, upper right panels) while the differences between \ce{HCN} and \ce{NH3} abundances have relatively subtle effect on transmission spectrum, highlighting the sensitivity of observations to \ce{CH4} in particular, at least for WASP-96b. \ce{CO2}, while exhibiting quenching, especially in the $10\times$ solar metallicity case, does not have significant differences between the simulations with and without the escape condition, and thus does not impact the transmission spectrum, unlike with \ce{CH4}. The differences between the transmission spectra for the UM/Venot simulations with full integration and the {\sc minichem} simulations are smaller, with the spectra differing by only 45, 42, and 30 ppm, for $0.1\times$, $1\times$, and $10\times$ solar metallicity, respectively, with the differences occurring primarily around \ce{CH4} features although a small disagreement in the \ce{CO2} feature at 4.3~$\mu$m occurs due to differences in the morning \ce{CO2} quenching between the UM/Venot and {\sc minichem} simulations (see Appendix \ref{Appendix:Additional}).  These comparisons occur without the inclusion of an offset applied to the transmission spectra; however, as these differences arise in part from both temperature differences, the overall differences can be reduced by 10, 8, and 3 ppm if an offset is applied. The differences in the predicted transmission spectra shown in Figure \ref{Fig:transmission} remain less than could be expected from other model choices \citep[see, e.g.,][]{steinrueck_2025}, and the use of the escape condition provides a larger impact on the synthetic observations than the choice of network.

\subsection{Impact on Emission}

\begin{figure*}
	\includegraphics[]{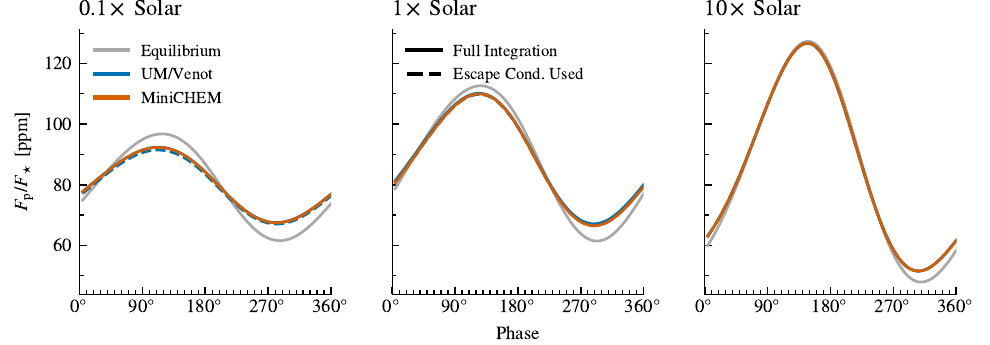}
    \caption{The integrated phase curve between 1.5 and 5 microns for each of the metallicities investigated.   In most cases, the chemical kinetics simulations are indistinguishable from one another, with the equilibrium chemistry simulations exhibiting the largest differences in emission.  The exception is for the UM/Venot 0.1$\times$ solar metallicity simulation that uses the escape condition where a larger offset from the other chemical kinetics simulations is observed. }
    \label{Fig:phasecurve}
\end{figure*}

We similarly investigate the impact of the chemical networks on the planetary emission, and we focus on the wavelengths between 1.5 and 5 microns, due to the influence of \ce{CH4} at  these wavelengths.    The integrated phase curve is shown in Figure \ref{Fig:phasecurve}.  While noticeable differences exist between the equilibrium cases and the chemical kinetics cases, the differences remain less than 7 ppm.  The differences between the individual chemical schemes is less than 1 ppm in all cases. Narrower bandpasses will, of course, result in larger differences.

\begin{figure*}
	\includegraphics[]{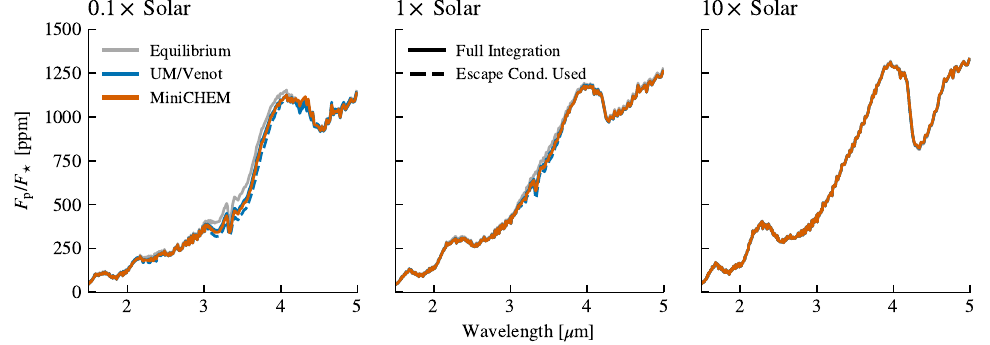}
    \caption{The ratio of planetary to stellar emission at a phase of 128\degree, roughly corresponding to the peak in the phase curve. }
    \label{Fig:peak_emission}
\end{figure*}

Examining the emission at a phase of 128\degree, roughly corresponding to the phase curve peaks, as shown in Figure \ref{Fig:peak_emission}, the differences in emission occur primarily between 3 and 4.3 microns, with the largest differences between the equilibrium and UM/Venot full integration simulations being 80, 85, and 8 ppm for the $0.1\times$, $1\times$, and $10\times$ solar metallicity cases, respectively.\footnote{In terms of relative differences in the spectra, the largest differences are 15\%, 14\%, and 0.8\% for the $0.1\times$, $1\times$, and $10\times$ solar metallicity cases, respectively.}  The UM/Venot and {\sc minichem} chemical kinetics differ by less than 48, 24, and 4 ppm for the $0.1\times$, $1\times$, and $10\times$ solar metallicity cases, respectively, with the differences again occurring in the 3 to 4.3 micron wavelength range, although unlike with the comparison with the equilibrium simulation where there is a noticeable different across the whole 3 to 4.3 micron interval, the differences when comparing the two chemical kinetics simulations are more localised around specific wavelengths.

\subsection{Performance}

The primary motivating factor for {\sc minichem} is the reduced computational overhead associated with the reduced network. We examine the wall-time to run 1 Earth day in each simulation, although as the calls to the chemistry solvers were not timed separately, the reported run times include all aspects of the computation.

Simulations were performed on the Max Planck Society's Viper high performance computer, each using 256 cores across 2 nodes. The time to simulate one Earth day for each of the simulations is shown in Table \ref{Tbl:SimTime}.

For a point of comparison of a simulation without any computationally intensive chemistry, we simulated the solar metallicity case for 100 Earth days using the analytic equilibrium chemistry scheme of \citet[][hereafter UM/BS99]{burrows_1999} which has been a default chemistry scheme in the  UM for hot Jupiters \citep{amundsen_2014a}, and which, compared to the schemes investigated here, has negligible computational overhead.  Although the UM/BS99 scheme does not employ tracers as abundances are estimated analytically, we perform separate simulations using the UM/BS99 scheme with zero, 13, and 30 tracer fields.  Within the implementation of the scheme, these tracers remain unused; however, they provide an estimate of the computational overhead of advecting the tracer fields in the UM/Venot and {\sc minichem} schemes, separate from the choice of chemistry scheme.  We find that without tracers, the UM/BS99 simulation takes 421 seconds to simulate one Earth day, which includes the time to perform dynamical and radiative transfer updates.  The inclusion of 13 tracers to the UM/BS99 simulation, as required by {\sc minichem}, and 30 tracers, as required by UM/Venot, increase the required time to 435 and 510 seconds, respectively.

The {\sc minichem} simulations required 505-595 seconds per Earth day, indicating that the chemistry including the tracers, and thus the use of {\sc minichem}, represents a $\sim$20\% to 40\% increase in simulation time.   The 13 tracers on their own, require only a 3\% increase in simulation time, relatively small compared to the cost of the chemical update.  The UM/Venot simulations require 30 tracers which, from the UM/BS99 tests, represents a $\sim$ 20\% increase in computational cost before the chemical update is factored in.  When the chemical update is accounted for, the UM/Venot simulations performing the full integration require 635 to 680 s per Earth day, effectively a $\sim$ 50 to 60\% increase in computational time.  Including the cost of tracer advection, the cost of the chemistry in the UM/Venot simulations performing the full integration is $\sim 1.5\times$ to $2.5\times$ that of {\sc minichem}, assuming the cost of all other aspects of the simulation is 421 s per Earth day.   

The slowest simulations, counter-intuitively, were the UM/Venot simulations that used the escape condition. This is likely due to the frequency of checks of the escape condition being sufficiently high that it has greater impact on the performance than any savings on performance due to the escape condition being triggered. This likely could be improved upon by reducing the frequency of checks, but this is beyond the scope of this investigation.

\begin{table}
\caption{Simulation time$^{a}$ for 1 Earth day, in seconds}
\label{Tbl:SimTime}
\begin{tabular}{lccc}
\hline
\hline
 & $0.1\times$ Solar & $1\times$ Solar & $10\times$ Solar \\
\hline
UM/BS99 (0 tracers) & - & 421 & -\\
UM/BS99 (13 tracers) & - & 435 & -\\
UM/BS99 (30 tracers) & - & 510 & -\\
{\sc minichem} & 505 & 529 & 595 \\
UM/Venot (Escape Cond.) & 663 & 695 & 721 \\
UM/Venot (Full Integration) & 635 & 660 & 680 \\
\hline
\multicolumn{4}{l}{$^\mathrm{a}$  {\footnotesize The quoted simulation time includes all aspects of the simulation,}} \\
\multicolumn{4}{l}{{\footnotesize including dynamics, radiative transfer, chemistry, and input/output.}}
\end{tabular}
\end{table}


\section{Conclusions}
\label{sec:conclusions}

We have benchmarked the {\sc minichem} chemical network against the \citetalias{venot_2019} network used with the UM's chemical kinetics solver, for three of the metallicities included in the {\sc minichem} repository and using the hot Jupiter WASP-96b as a test case. We have found excellent agreement in the abundances of the opacity sources tested, with the largest disagreements being with the quenching behaviour of \ce{HCN} and \ce{NH3}. These can likely be traced back to the absence of \ce{CH2NH} in the \citetalias{venot_2019} network and differing rates for the \ce{NH2 + NH2 -> N2H2 + H2} and \ce{NH3 + NH2 -> N2H3 + H2} reactions.  Neither network consistently has the more up-to-date reaction rates, making a recommendation difficult, but the use of the \citet{dean_1984} estimate for the second reaction and it's closer agreement to the \citet{marshall_2023} upper limit lead us to tentatively prefer the {\sc minichem} \ce{NH3} abundances over the UM/Venot abundances.  The use of the \citet{dean_1984} rate also increases the quenched \ce{HCN} abundance in the {\sc minichem} simulations, and the lack of \ce{CH2NH} in the \citetalias{venot_2019} network similarly leads us to prefer the {\sc minichem} \ce{HCN} abundances.  While we have compared the networks with the aspects relevant to GCM modelling in mind, this does not constitute a comprehensive analyses of all potential differences, and the \citet{venot_2012} network on which \citetalias{venot_2019} is based has since been updated by \citet{venot_2020} and \citet{veillet_2024}, although neither of these more recent networks yet offers a reduced network to be used in a GCM.   We also note that while we have found differences, the impact on the transmission spectra was found to be less than $\sim 45$ ppm, and in most parts of the spectrum significantly less,  which remains smaller than the impact of other modelling choices \citep[e.g.,][]{steinrueck_2025}.

With these differences in mind, {\sc minichem} offers noticeable increase in speed over the UM/Venot implementation, with the chemistry step likely being faster by a factor of 1.5 to 2.5, although we did not time the chemistry step explicitly in isolation in the benchmark. We note that while we have benchmarked this for a single target, {\sc minichem} offers rates covering a broad range of temperatures, and thoroughly studying planets across this parameter space would be computationally prohibitive. We have also not investigated the performance of {\sc minichem}, which currently only offers tables for solar carbon-to-oxygen ratios, when used for atmospheres with non-solar carbon-to-oxygen ratios.

The largest differences occurred not between the UM/Venot and {\sc minichem} simulations but instead during the tests of the escape condition in the UM chemical kinetics solver, which we found to result in an artificial increase in the quench pressure. We did not test tightening the tolerances of the escape condition, although we expect that this would bring the results into agreement at the cost of increased computational expense. As the use of the escape condition, as implemented, was found it ultimately slow the simulation, it is unclear whether an optimized version of the escape condition would provide significant benefit. It is likely prudent to avoid the use of this escape condition going forward, until its behaviour can be better quantified and any concerns assuaged. 

We end by stressing the importance of continual validation of methods and codes through intercomparison. While comparison against, for example, previous iterations of the same model can build confidence in the consistency of a code or model during development, comparison against other models provides an opportunity to understand the importance of differing underlying assumptions and to provide further context and confidence in the resulting science, as well as an opportunity to debug code. We highlight as being especially relevant to the work presented here the PIE ({Photochemical model Intercomparison for Exoplanet science}; Harmon et al., in prep), COD ACCRA ({Comparing One-DimentionAl Climates of Convective Radiative Atmosphere}; Chaverot, et al., in prep), as well as the ongoing International Space Science Institute (ISSI) 3D chemical kinetics intercomparison (ISSI \#23-597; PI: S-.M-. Tsai), all three of which are part of the CUISINES project \citep[Climates Using Interactive Suites of Intercomparisons  Nested for Exoplanet Studies;][]{sohl_2024}.

\section*{Acknowledgements}

The authors would like to thank the anonymous referee for their constructive comments. The authors would also like to thank Olivia Venot for providing the thermochemical data for her 2012 and 2019 networks. DAC is supported by the Max Planck Society. This research was also supported by a UK Research and Innovation (UKRI) Future Leaders Fellowship MR/T040866/1, and partly supported by the Leverhulme Trust through a research project grant RPG-2020-82 alongside a Science and Technology Facilities Council (STFC) (STFC) Small Award ST/Y00261X/1. E.K.H. Lee is supported by the CSH through the Bernoulli Fellowship. S-.M. T. is supported by the National Science and Technology Council (grants 114-2112-M-001-065-MY3).

The analysis of the simulation data made use of the following {\sc python} packages: {\sc aeolus} \citep{sergeev_2024}, {\sc iris} \citep{hattersley_2023}, {\sc matplotlib} \citep{hunter_2007}, and {\sc numpy} \citep{harris_2020}.

\section*{Data Availability}

The simulation data are available for download from the Zenodo online repository at \href{https://doi.org/10.5281/zenodo.19592855}{doi.org/10.5281/zenodo.19592855}. For the purpose of open access, the authors have applied a Creative Commons Attribution (CC BY) licence to any Author Accepted Manuscript version arising.



\bibliographystyle{mnras}
\bibliography{references} 

@phdthesis{drummond_2017,
	type = {Ph.{D}. thesis},
	title = {The chemistry of hot exoplanet atmospheres: {Developing} and applying chemistry schemes in {1D} and {3D} models},
	shorttitle = {The chemistry of hot exoplanet atmospheres},
	url = {https://ui.adsabs.harvard.edu/abs/2017PhDT.......209D},
	urldate = {2026-03-07},
	author = {Drummond, Benjamin},
	month = jan,
	year = {2017},
	note = {ADS Bibcode: 2017PhDT.......209D},
}

@misc{steinrueck_2025,
	title = {Limb {Asymmetries} on {WASP}-39b: {A} {Multi}-{GCM} {Comparison} of {Chemistry}, {Clouds}, and {Hazes}},
	shorttitle = {Limb {Asymmetries} on {WASP}-39b},
	doi = {10.48550/arXiv.2509.21588},
	publisher = {arXiv},
	author = {Steinrueck, Maria E. and Savel, Arjun B. and Christie, Duncan A. and Carone, Ludmila and Tsai, Shang-Min and Akın, Can and Kennedy, Thomas D. and Kiefer, Sven and Lewis, David A. and Rauscher, Emily and Samra, Dominic and Zamyatina, Maria and Arnold, Kenneth and Baeyens, Robin and Gkouvelis, Leonardos and Haegele, David and Helling, Christiane and Mayne, Nathan J. and Powell, Diana and Roman, Michael T. and Beltz, Hayley and Espinoza, Néstor and Heng, Kevin and Iro, Nicolas and Kempton, Eliza M. -R. and Kreidberg, Laura and Kirk, James and Murphy, Matthew M. and Rackham, Benjamin V. and Tan, Xianyu},
	month = sep,
	year = {2025},
}

@techreport{burcat_2005,
	title = {Third {Millennium} {Ideal} {Gas} and {Condensed} {Phase} {Thermochemical} {Database} for {Combustion} with {Updates} from {Active} {Thermochemical} {Tables}},
	language = {en},
	institution = {Argonne National Laboratory},
	author = {Burcat, Alexander and Ruscic, Branko},
	year = {2005},
}

@article{manna_2023,
	title = {{NH3NO} interaction at low-temperatures: {An} experimental and modeling study},
	volume = {39},
	issn = {15407489},
	shorttitle = {{NH3NO} interaction at low-temperatures},
	url = {https://linkinghub.elsevier.com/retrieve/pii/S1540748922003741},
	doi = {10.1016/j.proci.2022.09.027},
	language = {en},
	number = {1},
	urldate = {2025-11-17},
	journal = {Proceedings of the Combustion Institute},
	author = {Manna, Maria Virginia and Sabia, Pino and Shrestha, Krishna P. and Seidel, Lars and Ragucci, Raffaele and Mauss, Fabian and De Joannon, Mara},
	year = {2023},
	pages = {775--784},
}

@article{abian_2021,
	title = {Study of the oxidation of ammonia in a flow reactor. {Experiments} and kinetic modeling simulation},
	volume = {300},
	issn = {0016-2361},
	url = {https://www.sciencedirect.com/science/article/pii/S0016236121008565},
	doi = {10.1016/j.fuel.2021.120979},
	urldate = {2025-11-17},
	journal = {Fuel},
	author = {Abián, M. and Benés, M. and Goñi, A. de and Muñoz, B. and Alzueta, M. U.},
	month = sep,
	year = {2021},
	pages = {120979},
}

@misc{mcbride_1994,
	title = {Coefficients for {Calculating} {Thermodynamic} and {Transport} {Properties} of {Individual} {Species}},
	language = {en},
	author = {McBride, J and Gordon, Sanford and Reno, Martin},
	year = {1994},
}

@article{moses_2011,
	title = {Disequilibrium {Carbon}, {Oxygen}, and {Nitrogen} {Chemistry} in the {Atmospheres} of {HD} 189733b and {HD} 209458b},
	volume = {737},
	issn = {0004-637X},
	url = {https://ui.adsabs.harvard.edu/abs/2011ApJ...737...15M},
	doi = {10.1088/0004-637X/737/1/15},
	urldate = {2025-05-29},
	journal = {The Astrophysical Journal},
	publisher = {IOP},
	author = {Moses, Julianne I. and Visscher, C. and Fortney, J. J. and Showman, A. P. and Lewis, N. K. and Griffith, C. A. and Klippenstein, S. J. and Shabram, M. and Friedson, A. J. and Marley, M. S. and Freedman, R. S.},
	month = aug,
	year = {2011},
	note = {ADS Bibcode: 2011ApJ...737...15M},
	pages = {15},
}

@article{coppens_2007,
	title = {The effects of composition on burning velocity and nitric oxide formation in laminar premixed flames of {CH4} + {H2} + {O2} + {N2}},
	volume = {149},
	issn = {0010-2180},
	url = {https://www.sciencedirect.com/science/article/pii/S0010218007000636},
	doi = {10.1016/j.combustflame.2007.02.004},
	number = {4},
	urldate = {2025-11-08},
	journal = {Combustion and Flame},
	author = {Coppens, F. H. V. and De Ruyck, J. and Konnov, A. A.},
	month = jun,
	year = {2007},
	pages = {409--417},
}

@article{dean_1984,
	title = {Kinetics of rich ammonia flames},
	volume = {16},
	copyright = {Copyright © 1984 John Wiley \& Sons, Inc.},
	issn = {1097-4601},
	url = {https://onlinelibrary.wiley.com/doi/abs/10.1002/kin.550160603},
	doi = {10.1002/kin.550160603},
	language = {en},
	number = {6},
	urldate = {2025-11-07},
	journal = {International Journal of Chemical Kinetics},
	author = {Dean, Anthony M. and Chou, Mau-Song and Stern, David},
	year = {1984},
	note = {\_eprint: https://onlinelibrary.wiley.com/doi/pdf/10.1002/kin.550160603},
	pages = {633--653},
}

@article{dove_1979,
	title = {A shock-tube study of ammonia pyrolysis},
	volume = {57},
	issn = {0008-4042},
	url = {https://cdnsciencepub.com/doi/10.1139/v79-112},
	doi = {10.1139/v79-112},
	number = {6},
	urldate = {2025-11-07},
	journal = {Canadian Journal of Chemistry},
	publisher = {NRC Research Press},
	author = {Dove, John E. and Nip, Wing S.},
	month = mar,
	year = {1979},
	pages = {689--701},
}

@article{konnov_2000,
	title = {Kinetic {Modeling} of the {Thermal} {Decomposition} of {Ammonia}},
	volume = {152},
	issn = {0010-2202},
	url = {https://doi.org/10.1080/00102200008952125},
	doi = {10.1080/00102200008952125},
	number = {1},
	urldate = {2025-11-05},
	journal = {Combustion Science and Technology},
	publisher = {Taylor \& Francis},
	author = {Konnov, A.A. and Ruyck, J. DE},
	month = mar,
	year = {2000},
	note = {\_eprint: https://doi.org/10.1080/00102200008952125},
	pages = {23--37},
}

@article{davidson_1990,
	title = {A pyrolysis mechanism for ammonia},
	volume = {22},
	issn = {1097-4601},
	url = {https://onlinelibrary.wiley.com/doi/abs/10.1002/kin.550220508},
	doi = {10.1002/kin.550220508},
	language = {en},
	number = {5},
	urldate = {2025-11-07},
	journal = {International Journal of Chemical Kinetics},
	author = {Davidson, D. F. and Kohse-Höinghaus, K. and Chang, A. Y. and Hanson, R. K.},
	year = {1990},
	note = {\_eprint: https://onlinelibrary.wiley.com/doi/pdf/10.1002/kin.550220508},
	pages = {513--535},
}

@article{marshall_2023,
	title = {Probing {High}-{Temperature} {Amine} {Chemistry}: {Is} the {Reaction} {NH}$_{\textrm{3}}$ + {NH}$_{\textrm{2}}$ ⇄ {N}$_{\textrm{2}}$ {H}$_{\textrm{3}}$ + {H}$_{\textrm{2}}$ {Important}?},
	volume = {127},
	copyright = {https://doi.org/10.15223/policy-029},
	issn = {1089-5639, 1520-5215},
	shorttitle = {Probing {High}-{Temperature} {Amine} {Chemistry}},
	url = {https://pubs.acs.org/doi/10.1021/acs.jpca.2c08921},
	doi = {10.1021/acs.jpca.2c08921},
	language = {en},
	number = {11},
	urldate = {2025-11-07},
	journal = {The Journal of Physical Chemistry A},
	author = {Marshall, Paul and Glarborg, Peter},
	month = mar,
	year = {2023},
	pages = {2601--2607},
}

@article{wang_2023,
	title = {{NASA} {Polynomial} representation of molecular specific heats},
	volume = {306},
	issn = {0022-4073},
	url = {https://www.sciencedirect.com/science/article/pii/S0022407323001358},
	doi = {10.1016/j.jqsrt.2023.108617},
	urldate = {2025-10-26},
	journal = {Journal of Quantitative Spectroscopy and Radiative Transfer},
	author = {Wang, Rong and Balciunaite, Ugne and Chen, Juncai and Yuan, Cheng and Owens, Alec and Tennyson, Jonathan},
	month = sep,
	year = {2023},
	pages = {108617},
}

@article{moses_2014,
	title = {Chemical kinetics on extrasolar planets},
	volume = {372},
	url = {https://royalsocietypublishing.org/doi/10.1098/rsta.2013.0073},
	doi = {10.1098/rsta.2013.0073},
	number = {2014},
	urldate = {2025-10-26},
	journal = {Philosophical Transactions of the Royal Society A: Mathematical, Physical and Engineering Sciences},
	publisher = {Royal Society},
	author = {Moses, Julianne I.},
	month = apr,
	year = {2014},
	pages = {20130073},
}

@article{morley_2017,
	title = {Forward and {Inverse} {Modeling} of the {Emission} and {Transmission} {Spectrum} of {GJ} 436b: {Investigating} {Metal} {Enrichment}, {Tidal} {Heating}, and {Clouds}},
	volume = {153},
	issn = {0004-6256},
	shorttitle = {Forward and {Inverse} {Modeling} of the {Emission} and {Transmission} {Spectrum} of {GJ} 436b},
	url = {https://ui.adsabs.harvard.edu/abs/2017AJ....153...86M/abstract},
	doi = {10.3847/1538-3881/153/2/86},
	language = {en},
	number = {2},
	urldate = {2025-10-24},
	journal = {The Astronomical Journal, Volume 153, Issue 2, article id. 86, {\textless}NUMPAGES{\textgreater}15{\textless}/NUMPAGES{\textgreater} pp. (2017).},
	author = {Morley, Caroline V. and Knutson, Heather and Line, Michael and Fortney, Jonathan J. and Thorngren, Daniel and Marley, Mark S. and Teal, Dillon and Lupu, Roxana},
	month = feb,
	year = {2017},
	pages = {86},
}

@article{agundez_2014,
	title = {The {Puzzling} {Chemical} {Composition} of {GJ} 436b's {Atmosphere}: {Influence} of {Tidal} {Heating} on the {Chemistry}},
	volume = {781},
	issn = {0004-637X},
	shorttitle = {The {Puzzling} {Chemical} {Composition} of {GJ} 436b's {Atmosphere}},
	url = {https://ui.adsabs.harvard.edu/abs/2014ApJ...781...68A},
	doi = {10.1088/0004-637X/781/2/68},
	urldate = {2025-10-24},
	journal = {The Astrophysical Journal},
	publisher = {IOP},
	author = {Agúndez, Marcelino and Venot, Olivia and Selsis, Franck and Iro, Nicolas},
	month = feb,
	year = {2014},
	note = {ADS Bibcode: 2014ApJ...781...68A},
	pages = {68},
}

@article{ridgway_2023,
	title = {{3D} modelling of the impact of stellar activity on tidally locked terrestrial exoplanets: atmospheric composition and habitability},
	volume = {518},
	issn = {0035-8711},
	shorttitle = {{3D} modelling of the impact of stellar activity on tidally locked terrestrial exoplanets},
	url = {https://ui.adsabs.harvard.edu/abs/2023MNRAS.518.2472R},
	doi = {10.1093/mnras/stac3105},
	urldate = {2025-10-13},
	journal = {Monthly Notices of the Royal Astronomical Society},
	publisher = {OUP},
	author = {Ridgway, R. J. and Zamyatina, M. and Mayne, N. J. and Manners, J. and Lambert, F. H. and Braam, M. and Drummond, B. and Hébrard, E. and Palmer, P. I. and Kohary, K.},
	month = jan,
	year = {2023},
	note = {ADS Bibcode: 2023MNRAS.518.2472R},
	pages = {2472--2496},
}

@article{sergeev_2024,
	title = {Aeolus - a {Python} library for the analysis and visualisation of climate model output.},
	url = {https://ui.adsabs.harvard.edu/abs/2024zndo...5145603S},
	doi = {10.5281/zenodo.5145603},
	urldate = {2025-10-11},
	journal = {Zenodo},
	publisher = {Zenodo},
	author = {Sergeev, Denis E. and Zamyatina, Maria},
	month = nov,
	year = {2024},
	note = {ADS Bibcode: 2024zndo...5145603S},
}

@book{hairer_2010,
	address = {Berlin, Heidelberg},
	series = {Springer {Series} in {Computational} {Mathematics}},
	title = {Solving {Ordinary} {Differential} {Equations} {II}},
	volume = {14},
	copyright = {http://www.springer.com/tdm},
	isbn = {978-3-642-05220-0},
	doi = {10.1007/978-3-642-05221-7},
	publisher = {Springer},
	author = {Hairer, Ernst and Wanner, Gerhard},
	year = {2010},
}

@article{sohl_2024,
	title = {The {CUISINES} {Framework} for {Conducting} {Exoplanet} {Model} {Intercomparison} {Projects}, {Version} 1.0},
	volume = {5},
	issn = {2632-3338},
	url = {https://ui.adsabs.harvard.edu/abs/2024PSJ.....5..175S},
	doi = {10.3847/PSJ/ad5830},
	urldate = {2025-10-06},
	journal = {The Planetary Science Journal},
	publisher = {IOP},
	author = {Sohl, Linda E. and Fauchez, Thomas J. and Domagal-Goldman, Shawn and Christie, Duncan A. and Deitrick, Russell and Haqq-Misra, Jacob and Harman, C. E. and Iro, Nicolas and Mayne, Nathan J. and Tsigaridis, Kostas and Villanueva, Geronimo L. and Young, Amber V. and Chaverot, Guillaume},
	month = aug,
	year = {2024},
	note = {ADS Bibcode: 2024PSJ.....5..175S},
	pages = {175},
}

@article{klippenstein_2009,
	title = {Thermal {Decomposition} of {NH}$_{\textrm{2}}$ {OH} and {Subsequent} {Reactions}: {Ab} {Initio} {Transition} {State} {Theory} and {Reflected} {Shock} {Tube} {Experiments}},
	volume = {113},
	issn = {1089-5639, 1520-5215},
	shorttitle = {Thermal {Decomposition} of {NH}$_{\textrm{2}}$ {OH} and {Subsequent} {Reactions}},
	url = {https://pubs.acs.org/doi/10.1021/jp905454k},
	doi = {10.1021/jp905454k},
	language = {en},
	number = {38},
	urldate = {2025-09-22},
	journal = {The Journal of Physical Chemistry A},
	author = {Klippenstein, S. J. and Harding, L. B. and Ruscic, B. and Sivaramakrishnan, R. and Srinivasan, N. K. and Su, M.-C. and Michael, J. V.},
	month = sep,
	year = {2009},
	pages = {10241--10259},
}

@article{glarborg_2018,
	title = {Modeling nitrogen chemistry in combustion},
	volume = {67},
	issn = {03601285},
	url = {https://linkinghub.elsevier.com/retrieve/pii/S0360128517301600},
	doi = {10.1016/j.pecs.2018.01.002},
	language = {en},
	urldate = {2025-09-27},
	journal = {Progress in Energy and Combustion Science},
	author = {Glarborg, Peter and Miller, James A. and Ruscic, Branko and Klippenstein, Stephen J.},
	month = jul,
	year = {2018},
	pages = {31--68},
}

@misc{mcbride_2002,
	title = {{NASA} {Glenn} {Coefficients} for {Calculating} {Thermodynamic} {Properties} of {Individual} {Species}},
	author = {McBride, Bonnie and Zehe, Michael and Gordon, Sanford},
	month = sep,
	year = {2002},
}

@book{chase_1998,
	address = {Washington, D.C},
	edition = {4th ed},
	series = {Journal of physical and chemical reference data},
	title = {{NIST}-{JANAF} thermochemical tables},
	isbn = {978-1-56396-831-0 978-1-56396-819-8 978-1-56396-820-4},
	language = {eng},
	number = {9},
	publisher = {American chemical society},
	author = {Chase, Malcolm},
	collaborator = {{National institute of standards and technology}},
	year = {1998},
}

@article{tsai_2018,
	title = {Toward {Consistent} {Modeling} of {Atmospheric} {Chemistry} and {Dynamics} in {Exoplanets}: {Validation} and {Generalization} of the {Chemical} {Relaxation} {Method}},
	volume = {862},
	issn = {0004-637X},
	shorttitle = {Toward {Consistent} {Modeling} of {Atmospheric} {Chemistry} and {Dynamics} in {Exoplanets}},
	url = {https://ui.adsabs.harvard.edu/abs/2018ApJ...862...31T},
	doi = {10.3847/1538-4357/aac834},
	urldate = {2025-08-09},
	journal = {The Astrophysical Journal},
	publisher = {IOP},
	author = {Tsai, Shang-Min and Kitzmann, Daniel and Lyons, James R. and Mendonça, João and Grimm, Simon L. and Heng, Kevin},
	month = jul,
	year = {2018},
	note = {ADS Bibcode: 2018ApJ...862...31T},
	pages = {31},
}

@article{mendonca_2018,
	title = {Three-dimensional {Circulation} {Driving} {Chemical} {Disequilibrium} in {WASP}-43b},
	volume = {869},
	issn = {0004-637X},
	url = {https://ui.adsabs.harvard.edu/abs/2018ApJ...869..107M},
	doi = {10.3847/1538-4357/aaed23},
	urldate = {2025-08-07},
	journal = {The Astrophysical Journal},
	publisher = {IOP},
	author = {Mendonça, João M. and Tsai, Shang-min and Malik, Matej and Grimm, Simon L. and Heng, Kevin},
	month = dec,
	year = {2018},
	note = {ADS Bibcode: 2018ApJ...869..107M},
	pages = {107},
}

@article{steinrueck_2019,
	title = {The {Effect} of {3D} {Transport}-induced {Disequilibrium} {Carbon} {Chemistry} on the {Atmospheric} {Structure}, {Phase} {Curves}, and {Emission} {Spectra} of {Hot} {Jupiter} {HD} 189733b},
	volume = {880},
	issn = {0004-637X},
	url = {https://ui.adsabs.harvard.edu/abs/2019ApJ...880...14S},
	doi = {10.3847/1538-4357/ab2598},
	urldate = {2025-08-07},
	journal = {The Astrophysical Journal},
	publisher = {IOP},
	author = {Steinrueck, Maria E. and Parmentier, Vivien and Showman, Adam P. and Lothringer, Joshua D. and Lupu, Roxana E.},
	month = jul,
	year = {2019},
	note = {ADS Bibcode: 2019ApJ...880...14S},
	pages = {14},
}

@article{veillet_2024,
	title = {An extensively validated {C}/{H}/{O}/{N} chemical network for hot exoplanet disequilibrium chemistry},
	volume = {682},
	copyright = {https://creativecommons.org/licenses/by/4.0},
	issn = {0004-6361, 1432-0746},
	url = {https://www.aanda.org/10.1051/0004-6361/202346680},
	doi = {10.1051/0004-6361/202346680},
	language = {en},
	urldate = {2025-06-01},
	journal = {Astronomy \& Astrophysics},
	author = {Veillet, R. and Venot, O. and Sirjean, B. and Bounaceur, R. and Glaude, P.-A. and Al-Refaie, A. and Hébrard, E.},
	month = feb,
	year = {2024},
	pages = {A52},
}

@article{tsai_2017,
	title = {{VULCAN}: {An} {Open}-source, {Validated} {Chemical} {Kinetics} {Python} {Code} for {Exoplanetary} {Atmospheres}},
	volume = {228},
	issn = {0067-0049, 1538-4365},
	shorttitle = {{VULCAN}},
	url = {https://iopscience.iop.org/article/10.3847/1538-4365/228/2/20},
	doi = {10.3847/1538-4365/228/2/20},
	language = {en},
	number = {2},
	urldate = {2025-06-06},
	journal = {The Astrophysical Journal Supplement Series},
	author = {Tsai, Shang-Min and Lyons, James R. and Grosheintz, Luc and Rimmer, Paul B. and Kitzmann, Daniel and Heng, Kevin},
	month = feb,
	year = {2017},
	pages = {20},
}

@article{venot_2012,
	title = {A chemical model for the atmosphere of hot {Jupiters}},
	volume = {546},
	issn = {0004-6361, 1432-0746},
	url = {http://www.aanda.org/10.1051/0004-6361/201219310},
	doi = {10.1051/0004-6361/201219310},
	language = {en},
	urldate = {2025-05-29},
	journal = {Astronomy \& Astrophysics},
	author = {Venot, O. and Hébrard, E. and Agúndez, M. and Dobrijevic, M. and Selsis, F. and Hersant, F. and Iro, N. and Bounaceur, R.},
	month = oct,
	year = {2012},
	pages = {A43},
}

@article{radica_2023,
	title = {Awesome {SOSS}: transmission spectroscopy of {WASP}-96b with {NIRISS}/{SOSS}},
	volume = {524},
	issn = {0035-8711},
	shorttitle = {Awesome {SOSS}},
	url = {https://ui.adsabs.harvard.edu/abs/2023MNRAS.524..835R},
	doi = {10.1093/mnras/stad1762},
	urldate = {2025-05-26},
	journal = {Monthly Notices of the Royal Astronomical Society},
	publisher = {OUP},
	author = {Radica, Michael and Welbanks, Luis and Espinoza, Néstor and Taylor, Jake and Coulombe, Louis-Philippe and Feinstein, Adina D. and Goyal, Jayesh and Scarsdale, Nicholas and Albert, Loïc and Baghel, Priyanka and Bean, Jacob L. and Blecic, Jasmina and Lafrenière, David and MacDonald, Ryan J. and Zamyatina, Maria and Allart1, Romain and Artigau, Étienne and Batalha, Natasha E. and Cook, Neil James and Cowan, Nicolas B. and Dang, Lisa and Doyon, René and Fournier-Tondreau, Marylou and Johnstone, Doug and Line, Michael R. and Moran, Sarah E. and Mukherjee, Sagnick and Pelletier, Stefan and Roy, Pierre-Alexis and Talens, Geert Jan and Filippazzo, Joseph and Pontoppidan, Klaus and Volk, Kevin},
	month = sep,
	year = {2023},
	note = {ADS Bibcode: 2023MNRAS.524..835R},
	pages = {835--856},
}

@article{hellier_2014,
	title = {Transiting hot {Jupiters} from {WASP}-{South}, {Euler} and {TRAPPIST}: {WASP}-95b to {WASP}-101b},
	volume = {440},
	issn = {0035-8711},
	shorttitle = {Transiting hot {Jupiters} from {WASP}-{South}, {Euler} and {TRAPPIST}},
	url = {https://ui.adsabs.harvard.edu/abs/2014MNRAS.440.1982H},
	doi = {10.1093/mnras/stu410},
	urldate = {2025-05-26},
	journal = {Monthly Notices of the Royal Astronomical Society},
	publisher = {OUP},
	author = {Hellier, Coel and Anderson, D. R. and Collier Cameron, A. and Delrez, L. and Gillon, M. and Jehin, E. and Lendl, M. and Maxted, P. F. L. and Pepe, F. and Pollacco, D. and Queloz, D. and Ségransan, D. and Smalley, B. and Smith, A. M. S. and Southworth, J. and Triaud, A. H. M. J. and Udry, S. and West, R. G.},
	month = may,
	year = {2014},
	note = {ADS Bibcode: 2014MNRAS.440.1982H},
	pages = {1982--1992},
}

@article{hindmarsh_2019,
	title = {{ODEPACK}: {Ordinary} differential equation solver library},
	shorttitle = {{ODEPACK}},
	url = {https://ui.adsabs.harvard.edu/abs/2019ascl.soft05021H},
	urldate = {2025-05-26},
	journal = {Astrophysics Source Code Library},
	author = {Hindmarsh, Alan C.},
	month = may,
	year = {2019},
	note = {ADS Bibcode: 2019ascl.soft05021H},
	pages = {ascl:1905.021},
}

@article{taylor_2023,
	title = {Awesome {SOSS}: atmospheric characterization of {WASP}-96 b using the {JWST} early release observations},
	volume = {524},
	copyright = {https://creativecommons.org/licenses/by/4.0/},
	issn = {0035-8711, 1365-2966},
	shorttitle = {Awesome {SOSS}},
	url = {https://academic.oup.com/mnras/article/524/1/817/7177537},
	doi = {10.1093/mnras/stad1547},
	language = {en},
	number = {1},
	urldate = {2025-05-12},
	journal = {Monthly Notices of the Royal Astronomical Society},
	author = {Taylor, Jake and Radica, Michael and Welbanks, Luis and MacDonald, Ryan J and Blecic, Jasmina and Zamyatina, Maria and Roth, Alexander and Bean, Jacob L and Parmentier, Vivien and Coulombe, Louis-Philippe and Feinstein, Adina D and Espinoza, Néstor and Benneke, Björn and Lafrenière, David and Doyon, René and Ahrer, Eva-Maria},
	month = jul,
	year = {2023},
	pages = {817--834},
}

@article{sing_2024,
	title = {A warm {Neptune}’s methane reveals core mass and vigorous atmospheric mixing},
	volume = {630},
	copyright = {2024 The Author(s)},
	issn = {1476-4687},
	url = {https://www.nature.com/articles/s41586-024-07395-z},
	doi = {10.1038/s41586-024-07395-z},
	language = {en},
	number = {8018},
	urldate = {2025-01-20},
	journal = {Nature},
	publisher = {Nature Publishing Group},
	author = {Sing, David K. and Rustamkulov, Zafar and Thorngren, Daniel P. and Barstow, Joanna K. and Tremblin, Pascal and Alves de Oliveira, Catarina and Beck, Tracy L. and Birkmann, Stephan M. and Challener, Ryan C. and Crouzet, Nicolas and Espinoza, Néstor and Ferruit, Pierre and Giardino, Giovanna and Gressier, Amélie and Lee, Elspeth K. H. and Lewis, Nikole K. and Maiolino, Roberto and Manjavacas, Elena and Rauscher, Bernard J. and Sirianni, Marco and Valenti, Jeff A.},
	month = jun,
	year = {2024},
	pages = {831--835},
}

@article{welbanks_2024,
	title = {A high internal heat flux and large core in a warm {Neptune} exoplanet},
	volume = {630},
	copyright = {2024 The Author(s), under exclusive licence to Springer Nature Limited},
	issn = {1476-4687},
	url = {https://www.nature.com/articles/s41586-024-07514-w},
	doi = {10.1038/s41586-024-07514-w},
	language = {en},
	number = {8018},
	urldate = {2025-01-20},
	journal = {Nature},
	publisher = {Nature Publishing Group},
	author = {Welbanks, Luis and Bell, Taylor J. and Beatty, Thomas G. and Line, Michael R. and Ohno, Kazumasa and Fortney, Jonathan J. and Schlawin, Everett and Greene, Thomas P. and Rauscher, Emily and McGill, Peter and Murphy, Matthew and Parmentier, Vivien and Tang, Yao and Edelman, Isaac and Mukherjee, Sagnick and Wiser, Lindsey S. and Lagage, Pierre-Olivier and Dyrek, Achrène and Arnold, Kenneth E.},
	month = jun,
	year = {2024},
	pages = {836--840},
}

@article{christie_2024,
	title = {Longitudinal filtering, sponge layers, and equatorial jet formation in a general circulation model of gaseous exoplanets},
	volume = {532},
	copyright = {https://creativecommons.org/licenses/by/4.0/},
	issn = {0035-8711, 1365-2966},
	url = {https://academic.oup.com/mnras/article/532/3/3001/7689768},
	doi = {10.1093/mnras/stae1408},
	language = {en},
	number = {3},
	urldate = {2024-08-15},
	journal = {Monthly Notices of the Royal Astronomical Society},
	author = {Christie, D A and Mayne, N J and Zamyatina, M and Baskett, H and Evans-Soma, T M and Wood, N and Kohary, K},
	month = jul,
	year = {2024},
	pages = {3001--3019},
}

@article{harris_2020,
	title = {Array programming with {NumPy}},
	volume = {585},
	copyright = {2020 The Author(s)},
	issn = {1476-4687},
	url = {https://www.nature.com/articles/s41586-020-2649-2},
	doi = {10.1038/s41586-020-2649-2},
	language = {en},
	number = {7825},
	urldate = {2024-04-15},
	journal = {Nature},
	publisher = {Nature Publishing Group},
	author = {Harris, Charles R. and Millman, K. Jarrod and van der Walt, Stéfan J. and Gommers, Ralf and Virtanen, Pauli and Cournapeau, David and Wieser, Eric and Taylor, Julian and Berg, Sebastian and Smith, Nathaniel J. and Kern, Robert and Picus, Matti and Hoyer, Stephan and van Kerkwijk, Marten H. and Brett, Matthew and Haldane, Allan and del Río, Jaime Fernández and Wiebe, Mark and Peterson, Pearu and Gérard-Marchant, Pierre and Sheppard, Kevin and Reddy, Tyler and Weckesser, Warren and Abbasi, Hameer and Gohlke, Christoph and Oliphant, Travis E.},
	month = sep,
	year = {2020},
	pages = {357--362},
}

@article{hunter_2007,
	title = {Matplotlib: {A} {2D} {Graphics} {Environment}},
	volume = {9},
	copyright = {https://ieeexplore.ieee.org/Xplorehelp/downloads/license-information/IEEE.html},
	issn = {1521-9615},
	shorttitle = {Matplotlib},
	url = {http://ieeexplore.ieee.org/document/4160265/},
	doi = {10.1109/MCSE.2007.55},
	language = {en},
	number = {3},
	urldate = {2024-04-15},
	journal = {Computing in Science \& Engineering},
	author = {Hunter, John D.},
	year = {2007},
	pages = {90--95},
}

@article{hattersley_2023,
	title = {{SciTools}/iris: v3.7.0},
	shorttitle = {{SciTools}/iris},
	url = {https://ui.adsabs.harvard.edu/abs/2023zndo...8305232H},
	doi = {10.5281/zenodo.8305232},
	urldate = {2024-04-15},
	journal = {Zenodo},
	author = {Hattersley, Richard and Little, Bill and Peglar, Patrick and Elson, Phil and Campbell, Ed and Killick, Peter and Blay, Byron and Sales De Andrade, Elliott and {Lbdreyer} and Dawson, Andrew and Yeo, Martin and Comer, Ruth and Bosley, Corinne and Kirkham, Daniel and {Tkknight} and {Stephenworsley} and Benfold, Will and {Kwilliams-Mo} and {Tv3141} and {Filipe} and {Elias} and {Gm-S} and Leuprecht, Armin and Hoyer, Stephan and Robinson, Niall and Penn, James},
	month = aug,
	year = {2023},
	note = {ADS Bibcode: 2023zndo...8305232H},
}

@article{zamyatina_2023,
	title = {Observability of signatures of transport-induced chemistry in clear atmospheres of hot gas giant exoplanets},
	volume = {519},
	issn = {0035-8711, 1365-2966},
	url = {https://academic.oup.com/mnras/article/519/2/3129/6845749},
	doi = {10.1093/mnras/stac3432},
	language = {en},
	number = {2},
	urldate = {2023-01-18},
	journal = {Monthly Notices of the Royal Astronomical Society},
	author = {Zamyatina, Maria and Hébrard, Eric and Drummond, Benjamin and Mayne, Nathan J and Manners, James and Christie, Duncan A and Tremblin, Pascal and Sing, David K and Kohary, Krisztian},
	month = feb,
	year = {2023},
	pages = {3129--3153},
}

@article{zamyatina_2024,
	title = {Quenching-driven equatorial depletion and limb asymmetries in hot {Jupiter} atmospheres: {WASP}-96b example},
	volume = {529},
	issn = {0035-8711, 1365-2966},
	shorttitle = {Quenching-driven equatorial depletion and limb asymmetries in hot {Jupiter} atmospheres},
	url = {https://academic.oup.com/mnras/article/529/2/1776/7615449},
	doi = {10.1093/mnras/stae600},
	language = {en},
	number = {2},
	urldate = {2024-03-21},
	journal = {Monthly Notices of the Royal Astronomical Society},
	author = {Zamyatina, Maria and Christie, Duncan A and Hébrard, Eric and Mayne, Nathan J and Radica, Michael and Taylor, Jake and Baskett, Harry and Moore, Ben and Lils, Craig and Sergeev, Denis E and Ahrer, Eva-Maria and Manners, James and Kohary, Krisztian and Feinstein, Adina D},
	month = mar,
	year = {2024},
	pages = {1776--1801},
}

@article{sainsbury-martinez_2019,
	title = {Idealised simulations of the deep atmosphere of hot {Jupiters}: {Deep}, hot adiabats as a robust solution to the radius inflation problem},
	volume = {632},
	issn = {0004-6361, 1432-0746},
	shorttitle = {Idealised simulations of the deep atmosphere of hot {Jupiters}},
	url = {https://www.aanda.org/10.1051/0004-6361/201936445},
	doi = {10.1051/0004-6361/201936445},
	language = {en},
	urldate = {2024-02-08},
	journal = {Astronomy \& Astrophysics},
	author = {Sainsbury-Martinez, F. and Wang, P. and Fromang, S. and Tremblin, P. and Dubos, T. and Meurdesoif, Y. and Spiga, A. and Leconte, J. and Baraffe, I. and Chabrier, G. and Mayne, N. and Drummond, B. and Debras, F.},
	month = dec,
	year = {2019},
	pages = {A114},
}

@article{asplund_2009,
	title = {The {Chemical} {Composition} of the {Sun}},
	volume = {47},
	issn = {0066-4146, 1545-4282},
	url = {https://www.annualreviews.org/doi/10.1146/annurev.astro.46.060407.145222},
	doi = {10.1146/annurev.astro.46.060407.145222},
	language = {en},
	number = {1},
	urldate = {2024-01-28},
	journal = {Annual Review of Astronomy and Astrophysics},
	author = {Asplund, Martin and Grevesse, Nicolas and Sauval, A. Jacques and Scott, Pat},
	month = sep,
	year = {2009},
	pages = {481--522},
}

@article{caffau_2011,
	title = {Solar {Chemical} {Abundances} {Determined} with a {CO5BOLD} {3D} {Model} {Atmosphere}},
	volume = {268},
	issn = {0038-0938, 1573-093X},
	url = {http://link.springer.com/10.1007/s11207-010-9541-4},
	doi = {10.1007/s11207-010-9541-4},
	language = {en},
	number = {2},
	urldate = {2024-01-28},
	journal = {Solar Physics},
	author = {Caffau, E. and Ludwig, H.-G. and Steffen, M. and Freytag, B. and Bonifacio, P.},
	month = feb,
	year = {2011},
	pages = {255--269},
}

@article{tremblin_2017,
	title = {Advection of {Potential} {Temperature} in the {Atmosphere} of {Irradiated} {Exoplanets}: {A} {Robust} {Mechanism} to {Explain} {Radius} {Inflation}},
	volume = {841},
	issn = {1538-4357},
	shorttitle = {Advection of {Potential} {Temperature} in the {Atmosphere} of {Irradiated} {Exoplanets}},
	url = {https://iopscience.iop.org/article/10.3847/1538-4357/aa6e57},
	doi = {10.3847/1538-4357/aa6e57},
	language = {en},
	number = {1},
	urldate = {2023-08-21},
	journal = {The Astrophysical Journal},
	author = {Tremblin, P. and Chabrier, G. and Mayne, N. J. and Amundsen, D. S. and Baraffe, I. and Debras, F. and Drummond, B. and Manners, J. and Fromang, S.},
	month = may,
	year = {2017},
	pages = {30},
}

@article{venot_2020,
	title = {New chemical scheme for giant planet thermochemistry: {Update} of the methanol chemistry and new reduced chemical scheme},
	volume = {634},
	issn = {0004-6361, 1432-0746},
	shorttitle = {New chemical scheme for giant planet thermochemistry},
	url = {https://www.aanda.org/10.1051/0004-6361/201936697},
	doi = {10.1051/0004-6361/201936697},
	language = {en},
	urldate = {2023-07-13},
	journal = {Astronomy \& Astrophysics},
	author = {Venot, O. and Cavalié, T. and Bounaceur, R. and Tremblin, P. and Brouillard, L. and Lhoussaine Ben Brahim, R.},
	month = feb,
	year = {2020},
	pages = {A78},
}

@article{venot_2019,
	title = {Reduced chemical scheme for modelling warm to hot hydrogen-dominated atmospheres},
	volume = {624},
	issn = {0004-6361, 1432-0746},
	url = {https://www.aanda.org/10.1051/0004-6361/201834861},
	doi = {10.1051/0004-6361/201834861},
	language = {en},
	urldate = {2023-07-13},
	journal = {Astronomy \& Astrophysics},
	author = {Venot, O. and Bounaceur, R. and Dobrijevic, M. and Hébrard, E. and Cavalié, T. and Tremblin, P. and Drummond, B. and Charnay, B.},
	month = apr,
	year = {2019},
	pages = {A58},
}

@article{drummond_2020,
	title = {Implications of three-dimensional chemical transport in hot {Jupiter} atmospheres: {Results} from a consistently coupled chemistry-radiation-hydrodynamics model},
	volume = {636},
	issn = {0004-6361},
	shorttitle = {Implications of three-dimensional chemical transport in hot {Jupiter} atmospheres},
	url = {https://ui.adsabs.harvard.edu/abs/2020A&A...636A..68D},
	doi = {10.1051/0004-6361/201937153},
	urldate = {2023-05-23},
	journal = {Astronomy and Astrophysics},
	author = {Drummond, Benjamin and Hébrard, Eric and Mayne, Nathan J. and Venot, Olivia and Ridgway, Robert J. and Changeat, Quentin and Tsai, Shang-Min and Manners, James and Tremblin, Pascal and Abraham, Nathan Luke and Sing, David and Kohary, Krisztian},
	month = apr,
	year = {2020},
	note = {ADS Bibcode: 2020A\&A...636A..68D},
	pages = {A68},
}

@article{tan_2019,
	title = {The {Atmospheric} {Circulation} of {Ultra}-hot {Jupiters}},
	volume = {886},
	issn = {1538-4357},
	url = {https://iopscience.iop.org/article/10.3847/1538-4357/ab4a76},
	doi = {10.3847/1538-4357/ab4a76},
	language = {en},
	number = {1},
	urldate = {2023-05-23},
	journal = {The Astrophysical Journal},
	author = {Tan, Xianyu and Komacek, Thaddeus D.},
	month = nov,
	year = {2019},
	pages = {26},
}

@article{cooper_2006,
	title = {Dynamics and {Disequilibrium} {Carbon} {Chemistry} in {Hot} {Jupiter} {Atmospheres}, with {Application} to {HD} 209458b},
	volume = {649},
	issn = {0004-637X, 1538-4357},
	url = {https://iopscience.iop.org/article/10.1086/506312},
	doi = {10.1086/506312},
	language = {en},
	number = {2},
	urldate = {2023-07-09},
	journal = {The Astrophysical Journal},
	author = {Cooper, Curtis S. and Showman, Adam P.},
	month = oct,
	year = {2006},
	pages = {1048--1063},
}

@article{lee_2023,
	title = {A mini-chemical scheme with net reactions for {3D} general circulation models: {II}. {3D} thermochemical modelling of {WASP}-39b and {HD} 189733b},
	volume = {672},
	issn = {0004-6361, 1432-0746},
	shorttitle = {A mini-chemical scheme with net reactions for {3D} general circulation models},
	url = {https://www.aanda.org/10.1051/0004-6361/202245473},
	doi = {10.1051/0004-6361/202245473},
	language = {en},
	urldate = {2023-07-09},
	journal = {Astronomy \& Astrophysics},
	author = {Lee, Elspeth K. H. and Tsai, Shang-Min and Hammond, Mark and Tan, Xianyu},
	month = apr,
	year = {2023},
	pages = {A110},
}

@article{braam_2022,
	title = {Lightning-induced chemistry on tidally-locked {Earth}-like exoplanets},
	volume = {517},
	issn = {0035-8711, 1365-2966},
	url = {https://academic.oup.com/mnras/article/517/2/2383/6717659},
	doi = {10.1093/mnras/stac2722},
	language = {en},
	number = {2},
	urldate = {2023-02-02},
	journal = {Monthly Notices of the Royal Astronomical Society},
	author = {Braam, Marrick and Palmer, Paul I and Decin, Leen and Ridgway, Robert J and Zamyatina, Maria and Mayne, Nathan J and Sergeev, Denis E and Abraham, N Luke},
	month = oct,
	year = {2022},
	pages = {2383--2402},
}

@article{tsai_2022,
	title = {A mini-chemical scheme with net reactions for {3D} general circulation models: {I}. {Thermochemical} kinetics},
	volume = {664},
	issn = {0004-6361, 1432-0746},
	shorttitle = {A mini-chemical scheme with net reactions for {3D} general circulation models},
	url = {https://www.aanda.org/10.1051/0004-6361/202142816},
	doi = {10.1051/0004-6361/202142816},
	language = {en},
	urldate = {2023-02-02},
	journal = {Astronomy \& Astrophysics},
	author = {Tsai, Shang-Min and Lee, Elspeth K. H. and Pierrehumbert, Raymond},
	month = aug,
	year = {2022},
	pages = {A82},
}

@article{christie_2022,
	title = {The impact of phase equilibrium cloud models on {GCM} simulations of {GJ} 1214b},
	volume = {517},
	issn = {0035-8711},
	url = {https://ui.adsabs.harvard.edu/abs/2022MNRAS.517.1407C},
	doi = {10.1093/mnras/stac2763},
	urldate = {2022-12-01},
	journal = {Monthly Notices of the Royal Astronomical Society},
	author = {Christie, D. A. and Mayne, N. J. and Gillard, R. M. and Manners, J. and Hébrard, E. and Lines, S. and Kohary, K.},
	month = nov,
	year = {2022},
	note = {ADS Bibcode: 2022MNRAS.517.1407C},
	pages = {1407--1421},
}

@article{jackson_2020,
	title = {The {Space} {Weather} {Atmosphere} {Models} and {Indices} ({SWAMI}) project: {Overview} and first results},
	volume = {10},
	issn = {2115-7251},
	shorttitle = {The {Space} {Weather} {Atmosphere} {Models} and {Indices} ({SWAMI}) project},
	url = {https://www.swsc-journal.org/10.1051/swsc/2020019},
	doi = {10.1051/swsc/2020019},
	language = {en},
	urldate = {2022-04-01},
	journal = {Journal of Space Weather and Space Climate},
	author = {Jackson, David R. and Bruinsma, Sean and Negrin, Sandra and Stolle, Claudia and Budd, Chris J. and Dominguez Gonzalez, Raul and Down, Emily and Griffin, Daniel J. and Griffith, Matthew J. and Kervalishvili, Guram and Lubián Arenillas, Daniel and Manners, James and Matzka, Jürgen and Shprits, Yuri Y. and Vasile, Ruggero and Zhelavskaya, Irina S.},
	year = {2020},
	pages = {18},
}

@article{drummond_2018,
	title = {The effect of metallicity on the atmospheres of exoplanets with fully coupled {3D} hydrodynamics, equilibrium chemistry, and radiative transfer},
	volume = {612},
	issn = {0004-6361, 1432-0746},
	url = {http://arxiv.org/abs/1801.01045},
	doi = {10.1051/0004-6361/201732010},
	language = {en},
	urldate = {2022-02-11},
	journal = {Astronomy \& Astrophysics},
	author = {Drummond, Benjamin and Mayne, N. J. and Baraffe, Isabelle and Tremblin, Pascal and Manners, James and Amundsen, David S. and Goyal, Jayesh and Acreman, Dave},
	month = apr,
	year = {2018},
	note = {arXiv: 1801.01045},
	pages = {A105},
}

@article{drummond_2016,
	title = {The effects of consistent chemical kinetics calculations on the pressure-temperature profiles and emission spectra of hot {Jupiters}},
	volume = {594},
	issn = {0004-6361, 1432-0746},
	url = {http://www.aanda.org/10.1051/0004-6361/201628799},
	doi = {10.1051/0004-6361/201628799},
	language = {en},
	urldate = {2022-02-03},
	journal = {Astronomy \& Astrophysics},
	author = {Drummond, B. and Tremblin, P. and Baraffe, I. and Amundsen, D. S. and Mayne, N. J. and Venot, O. and Goyal, J.},
	month = oct,
	year = {2016},
	pages = {A69},
}

@article{tremblin_2015,
	title = {{FINGERING} {CONVECTION} {AND} {CLOUDLESS} {MODELS} {FOR} {COOL} {BROWN} {DWARF} {ATMOSPHERES}},
	volume = {804},
	issn = {2041-8213},
	url = {https://iopscience.iop.org/article/10.1088/2041-8205/804/1/L17},
	doi = {10.1088/2041-8205/804/1/L17},
	language = {en},
	number = {1},
	urldate = {2022-02-03},
	journal = {The Astrophysical Journal},
	author = {Tremblin, P. and Amundsen, D. S. and Mourier, P. and Baraffe, I. and Chabrier, G. and Drummond, B. and Homeier, D. and Venot, O.},
	month = apr,
	year = {2015},
	pages = {L17},
}

@article{edwards_1996,
	title = {Studies with a flexible new radiation code. {I}: {Choosing} a configuration for a large-scale model},
	volume = {122},
	issn = {00359009, 1477870X},
	shorttitle = {Studies with a flexible new radiation code. {I}},
	url = {https://onlinelibrary.wiley.com/doi/10.1002/qj.49712253107},
	doi = {10.1002/qj.49712253107},
	language = {en},
	number = {531},
	urldate = {2022-02-03},
	journal = {Quarterly Journal of the Royal Meteorological Society},
	author = {Edwards, J. M. and Slingo, A.},
	month = apr,
	year = {1996},
	pages = {689--719},
}

@article{wood_2014,
	title = {An inherently mass-conserving semi-implicit semi-{Lagrangian} discretization of the deep-atmosphere global non-hydrostatic equations},
	volume = {140},
	issn = {00359009},
	url = {https://onlinelibrary.wiley.com/doi/10.1002/qj.2235},
	doi = {10.1002/qj.2235},
	language = {en},
	number = {682},
	urldate = {2021-10-04},
	journal = {Quarterly Journal of the Royal Meteorological Society},
	author = {Wood, Nigel and Staniforth, Andrew and White, Andy and Allen, Thomas and Diamantakis, Michail and Gross, Markus and Melvin, Thomas and Smith, Chris and Vosper, Simon and Zerroukat, Mohamed and Thuburn, John},
	month = jul,
	year = {2014},
	pages = {1505--1520},
}

@article{mayne_2014a,
	title = {The unified model, a fully-compressible, non-hydrostatic, deep atmosphere global circulation model, applied to hot {Jupiters}: {ENDGame} for a {HD} 209458b test case},
	volume = {561},
	issn = {0004-6361, 1432-0746},
	shorttitle = {The unified model, a fully-compressible, non-hydrostatic, deep atmosphere global circulation model, applied to hot {Jupiters}},
	url = {http://www.aanda.org/10.1051/0004-6361/201322174},
	doi = {10.1051/0004-6361/201322174},
	language = {en},
	urldate = {2021-09-20},
	journal = {Astronomy \& Astrophysics},
	author = {Mayne, Nathan J. and Baraffe, Isabelle and Acreman, David M. and Smith, Chris and Browning, Matthew K. and Amundsen, David Skålid and Wood, Nigel and Thuburn, John and Jackson, David R.},
	month = jan,
	year = {2014},
	pages = {A1},
}

@article{mayne_2019,
	title = {The {Limits} of the {Primitive} {Equations} of {Dynamics} for {Warm}, {Slowly} {Rotating} {Small} {Neptunes} and {Super} {Earths}},
	volume = {871},
	issn = {1538-4357},
	url = {https://iopscience.iop.org/article/10.3847/1538-4357/aaf6e9},
	doi = {10.3847/1538-4357/aaf6e9},
	language = {en},
	number = {1},
	urldate = {2021-09-20},
	journal = {The Astrophysical Journal},
	author = {Mayne, N. J. and Drummond, B. and Debras, F. and Jaupart, E. and Manners, J. and Boutle, I. A. and Baraffe, I. and Kohary, K.},
	month = jan,
	year = {2019},
	pages = {56},
}

@article{lee_2021,
	title = {Simulating gas giant exoplanet atmospheres with {\textless}span style="font-variant:small-caps;"{\textgreater}{Exo}-{FMS}{\textless}/span{\textgreater} : comparing semigrey, picket fence, and correlated- \textit{k} radiative-transfer schemes},
	volume = {506},
	issn = {0035-8711, 1365-2966},
	shorttitle = {Simulating gas giant exoplanet atmospheres with {\textless}span style="font-variant},
	url = {https://academic.oup.com/mnras/article/506/2/2695/6316122},
	doi = {10.1093/mnras/stab1851},
	language = {en},
	number = {2},
	urldate = {2021-09-20},
	journal = {Monthly Notices of the Royal Astronomical Society},
	author = {Lee, Elspeth K H and Parmentier, Vivien and Hammond, Mark and Grimm, Simon L and Kitzmann, Daniel and Tan, Xianyu and Tsai, Shang-Min and Pierrehumbert, Raymond T},
	month = jul,
	year = {2021},
	pages = {2695--2711},
}

@article{christie_2021,
	title = {The impact of mixing treatments on cloud modelling in {3D} simulations of hot {Jupiters}},
	volume = {506},
	issn = {0035-8711, 1365-2966},
	url = {https://academic.oup.com/mnras/article/506/3/4500/6322843},
	doi = {10.1093/mnras/stab2027},
	language = {en},
	number = {3},
	urldate = {2021-10-04},
	journal = {Monthly Notices of the Royal Astronomical Society},
	author = {Christie, D A and Mayne, N J and Lines, S and Parmentier, V and Manners, J and Boutle, I and Drummond, B and Mikal-Evans, T and Sing, D K and Kohary, K},
	month = aug,
	year = {2021},
	pages = {4500--4515},
}

@article{lines_2018b,
	title = {Exonephology: {Transmission} spectra from a {3D} simulated cloudy atmosphere of {HD209458b}},
	volume = {481},
	issn = {0035-8711, 1365-2966},
	shorttitle = {Exonephology},
	url = {http://arxiv.org/abs/1808.05887},
	doi = {10.1093/mnras/sty2275},
	language = {en},
	number = {1},
	urldate = {2020-12-07},
	journal = {Monthly Notices of the Royal Astronomical Society},
	author = {Lines, S. and Manners, J. and Mayne, N. J. and Goyal, J. and Carter, A. L. and Boutle, I. A. and Lee, E. K. H. and Helling, Ch and Drummond, B. and Acreman, D. M. and Sing, D. K.},
	month = nov,
	year = {2018},
	note = {arXiv: 1808.05887},
	pages = {194--205},
}

@article{burrows_1999,
	title = {Chemical {Equilibrium} {Abundances} in {Brown} {Dwarf} and {Extrasolar} {Giant} {Planet} {Atmospheres}},
	volume = {512},
	issn = {0004-637X, 1538-4357},
	url = {https://iopscience.iop.org/article/10.1086/306811},
	doi = {10.1086/306811},
	language = {en},
	number = {2},
	urldate = {2021-03-24},
	journal = {The Astrophysical Journal},
	author = {Burrows, Adam and Sharp, C. M.},
	month = feb,
	year = {1999},
	pages = {843--863},
}

@article{amundsen_2016,
	title = {The {UK} {Met} {Office} global circulation model with a sophisticated radiation scheme applied to the hot {Jupiter} {HD} 209458b},
	volume = {595},
	issn = {0004-6361, 1432-0746},
	url = {http://www.aanda.org/10.1051/0004-6361/201629183},
	doi = {10.1051/0004-6361/201629183},
	language = {en},
	urldate = {2021-03-23},
	journal = {Astronomy \& Astrophysics},
	author = {Amundsen, David S. and Mayne, Nathan J. and Baraffe, Isabelle and Manners, James and Tremblin, Pascal and Drummond, Benjamin and Smith, Chris and Acreman, David M. and Homeier, Derek},
	month = nov,
	year = {2016},
	pages = {A36},
}

@article{amundsen_2017,
	title = {Treatment of overlapping gaseous absorption with the correlated- \textit{k} method in hot {Jupiter} and brown dwarf atmosphere models},
	volume = {598},
	issn = {0004-6361, 1432-0746},
	url = {http://www.aanda.org/10.1051/0004-6361/201629322},
	doi = {10.1051/0004-6361/201629322},
	language = {en},
	urldate = {2021-02-24},
	journal = {Astronomy \& Astrophysics},
	author = {Amundsen, David S. and Tremblin, Pascal and Manners, James and Baraffe, Isabelle and Mayne, Nathan J.},
	month = feb,
	year = {2017},
	pages = {A97},
}

@article{amundsen_2014a,
	title = {Accuracy tests of radiation schemes used in hot {Jupiter} global circulation models},
	volume = {564},
	issn = {0004-6361, 1432-0746},
	url = {http://arxiv.org/abs/1402.0814},
	doi = {10.1051/0004-6361/201323169},
	language = {en},
	urldate = {2021-02-24},
	journal = {Astronomy \& Astrophysics},
	author = {Amundsen, David Skålid and Baraffe, Isabelle and Tremblin, Pascal and Manners, James and Hayek, Wolfgang and Mayne, N. J. and Acreman, David M.},
	month = apr,
	year = {2014},
	note = {arXiv: 1402.0814},
	pages = {A59},
}

@article{fortney_2020,
	title = {Beyond {Equilibrium} {Temperature}: {How} the {Atmosphere}/{Interior} {Connection} {Affects} the {Onset} of {Methane}, {Ammonia}, and {Clouds} in {Warm} {Transiting} {Giant} {Planets}},
	volume = {160},
	issn = {1538-3881},
	shorttitle = {Beyond {Equilibrium} {Temperature}},
	url = {http://arxiv.org/abs/2010.00146},
	doi = {10.3847/1538-3881/abc5bd},
	language = {en},
	number = {6},
	urldate = {2021-02-15},
	journal = {The Astronomical Journal},
	author = {Fortney, Jonathan J. and Visscher, Channon and Marley, Mark S. and Hood, Callie E. and Line, Michael R. and Thorngren, Daniel P. and Freedman, Richard S. and Lupu, Roxana},
	month = nov,
	year = {2020},
	note = {arXiv: 2010.00146},
	pages = {288},
}




\appendix


\section{Thermochemical Data and Methane Abundance}
\label{Appendix:Methane}
As the thermochemical data used in the \citetalias{venot_2019} and in {\sc minichem} differ at temperatures above 1000~K, we examine how these differences impact equilibrium \ce{ch4} abundances.  Using the equilibrium chemistry solver in {\sc atmo}, we solved for the equilibrium \ce{CH4} abundances on a grid of temperatures and pressures, for each of the metallicities investigated in this work.

\begin{figure*}
	\includegraphics[]{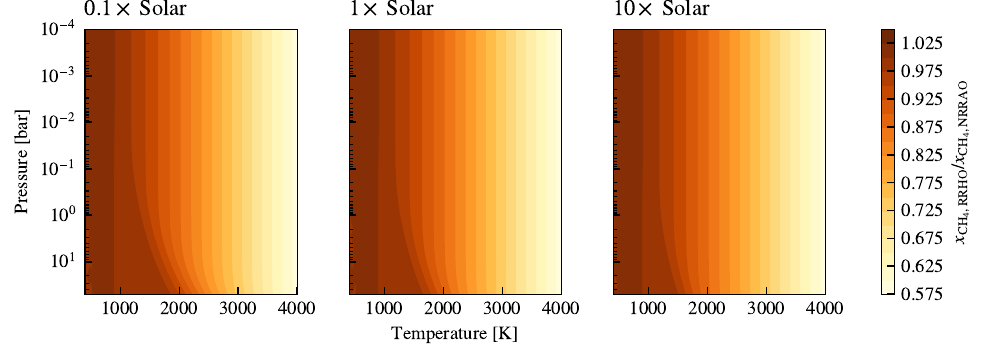}
    \caption{The ratio of equilibrium volume mixing ratios for \ce{CH4} calculated using the RRHO approximation versus using the NRRAO approximation, for each of the three metallicities investigated here.}
    \label{Fig:ch4_comp}
\end{figure*}

The results are shown in Figure \ref{Fig:ch4_comp}.  For the temperatures above 1000~K, the rigid rotation harmonic oscillator (RRHO) approximation results in an underestimation relative to the recommended non-rigid rotator anharmonic oscillator (NRRAO) approximation used in the \citet{mcbride_2002} fits. For temperatures found in the simulations in this work, the differences are on the order of a few percent, but as temperatures increase, the discrepancy increases, reaching in excess of a 40\% at temperatures of 4000~K.

\section{Modification to \ce{N2H2} and \ce{N2H3} Rates}
\label{Appendix:Rates}

\begin{figure*}
	\includegraphics[]{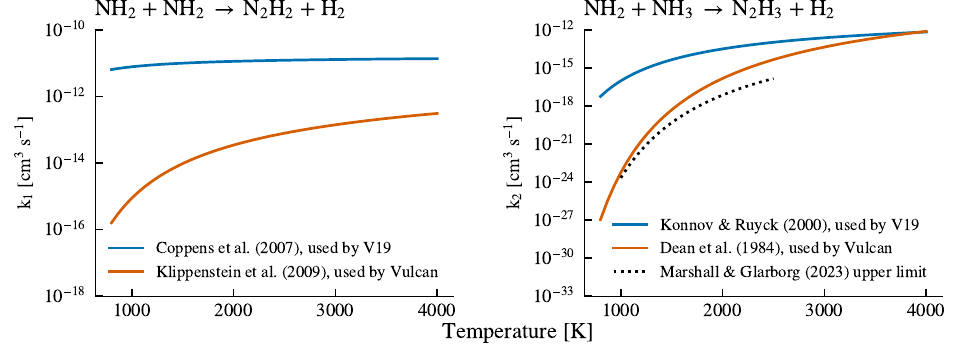}
    \caption{A comparison of the rates $k_1$(\ce{NH2 + NH2 -> N2H2 + H2}) and $k_2$(\ce{NH2 + NH3 -> N2H3 + H2}) examined in Appendix \ref{Appendix:Rates}.}
    \label{Fig:n2hx_rates}
\end{figure*}

\begin{figure*}
	\includegraphics[]{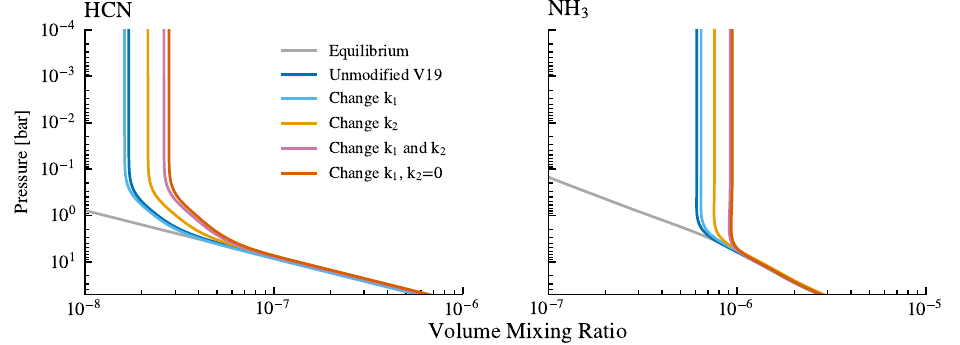}
    \caption{The results from 1D {\sc atmo} simulations using the original \citetalias{venot_2019} network as well as cases where individual rates are replaced with the rates used in {\sc vulcan}. The area-weighted abundances for \ce{HCN} (left panel) and \ce{NH3} (right panel) for solar metallicity simulations, each run for $10^9$ seconds. For reference, the equilibrium abundance profiles are shown in grey.}
    \label{Fig:atmo_test}
\end{figure*}

To understand to what extent reactions involving \ce{N2H_x} are impacting the differences in \ce{HCN} and \ce{NH3} abundances between UM/Venot and {\sc minichem} simulations, we look at the impact of changing two rates within the \citetalias{venot_2019} network such that they agree with the rate in used in {\sc vulcan}, on which {\sc minichem} is based.  A summary of the rates are found in Figure \ref{Fig:n2hx_rates}.

For the first reaction, \ce{NH2 + NH2 -> N2H2 + H2}, \citet{venot_2012,venot_2019} adopt the rate used in \citet{coppens_2007},

\begin{equation}
    k_{1,\mathrm{V19}} = 1.66\times 10^{-11}\exp\left(-754\,\mathrm{K}/T\right)\,\, \mathrm{cm^3\,s^{-1}},
\end{equation}
\noindent while {\sc vulcan} adopts the slightly more recent rate from \citet{klippenstein_2009},

\begin{equation}
    k_{1,\mathrm{vulcan}} = 9.717\times 10^{-14}\left(\frac{T}{300\,\mathrm{K}}\right)^{1.02}\exp\left(-5930\,\mathrm{K}/T\right)\,\,\mathrm{cm^3\,s^{-1}}.
\end{equation}

\noindent  This rate is included in other recent networks \citep[e.g.,][]{glarborg_2018,veillet_2024}, and we take it to be the more accurate of the two.

In the second reaction, \ce{NH2 + NH3 -> N2H3 + H2}, \citet{venot_2012,venot_2019} use the rate from \citet{konnov_2000}, 
\begin{equation}
    k_{2,\mathrm{V19}} = 2.88\times 10^{-12}\left(\frac{T}{300\,\mathrm{K}}\right)^{0.5}\exp\left(-10900\,\mathrm{K}/T\right)\,\, \mathrm{cm^3\,s^{-1}},
\end{equation}

\noindent which is the rate from \citet{dove_1979} scaled down by a factor of eight to improve agreement with the pyrolysis experiments of \citet{davidson_1990}.   {\sc vulcan}, following \citet{moses_2011},  instead uses the estimate from \citet{dean_1984},

\begin{equation}
    k_{2,\mathrm{vulcan}} = 4\times 10^{-9}\exp\left(-34200\,\mathrm{K}/T\right)\,\, \mathrm{cm^3\,s^{-1}}.
\end{equation}

\noindent The rate for this reaction is poorly constrained (see the discussion in \citealt{marshall_2023}), and while the rates $k_{2,\mathrm{vulcan}}$ is smaller than $ k_{2,\mathrm{V19}}$ by orders of magnitude, other chemical kinetic models opt to omit the reaction entirely \citep[e.g.,][]{glarborg_2018,veillet_2024} and express uncertainty as to whether it even occurs.  \citet{abian_2021}, on the other hand, adopt the original \citet{dove_1979} rate, unscaled, while \citet{manna_2023} use the \citet{dove_1979}, increased by a factor of two, in modelling experimental flow reactor data.  \citet{marshall_2023} provide an summary of the various rates used in the literature, and estimate an upper bound on the rate,

\begin{equation}
    k_{2,\mathrm{Marshall}} = 2.491\times 10^{-11}\exp\left(-30092\,\mathrm{K}/T\right)\,\, \mathrm{cm^3\,s^{-1}},
\end{equation}

\noindent for temperatures 1000-2500\,K, which, although an upper limit, aligns closer to the estimated rate of \citet[][see also Figure \ref{Fig:n2hx_rates}]{dean_1984}.   We note that this represents an unresolved tension in the literature, and as is expressed in \citet{marshall_2023}, either the reaction rate is in fact ``improbably high'' or there are unmodelled reaction pathways whose omission are compensated for through the increase in this reaction rate.   If the latter is the case, the compensating effect needed to reproduce experimental results in the laboratory may not have the same effect in modelling hot Jupiter atmospheres which have different compositions from experimental setups, and thus caution should be exercised.

\subsection{Tests with {\sc atmo}}

For a less computationally-costly method to investigate the impact of rates, we simulate one-dimensional solar metallicity atmospheres of WASP-96b, using the area-weighted average pressure-temperature profile from the equilibrium GCM simulation and a fixed $K_{zz} = 10^{10}\,\mathrm{cm^2\,s^{-1}}$.   In addition to simulations with equilibrium chemistry and with the \citetalias{venot_2019} network, we test four modifications to the \citetalias{venot_2019} model, one where $k_1$ is replaced with the {\sc vulcan} rate $k_{1,\mathrm{vulcan}}$, one where $k_2$ is replaced with the {\sc vulcan} rate $k_{2,\mathrm{vulcan}}$, one model where both changes are made, and one model where $k_1$ is replaced with the {\sc vulcan} rate and $k_2=0$ (i.e., removing the second reaction from the network).

The results are shown in Figure \ref{Fig:atmo_test}.  Replacing the rate for \ce{NH2 + NH2 -> N2H2 + H2} with the more recent \citet{klippenstein_2009} rate results in a small change in the quenched \ce{HCN} and \ce{NH3} abundances while replacing only the \ce{NH3 + NH2 -> N2H3 + H2} rate with the older \citet{dean_1984} estimate increases the quenched \ce{HCN} and \ce{NH3} abundances by 26\% and 24\%, respectively.   Changing both rates results in a larger increase of 54\% and 50\% for \ce{HCN} and \ce{NH3}, respectively, indicating the increased importance of the first reaction once the second reaction rate is changed.   Eliminating the second reaction entirely further increases the \ce{HCN} and \ce{NH3} abundances, although only moderately.

We conclude that the differences between the \citet{konnov_2000} and \citet{dean_1984} rates for the reaction \ce{NH3 + NH2 -> N2H3 + H2} are the primary source for the reduced quenched \ce{NH3} abundance, and these differences is partially responsible  for the smaller \ce{HCN} abundances, although there are reasons to believe other mechanisms, such as the inclusion of additional species, will further contribute to the observed disagreement.

\subsection{Test in the UM}

As the thermal structure and transport in the GCM simulations is more complicated than in a 1D model with fixed $K_{zz}$, we perform a test with a solar metallicity atmosphere and the UM/Venot chemical kinetics setup, except both reactions in the previous section have their rates replaced with the rates used in {\sc vulcan}, as was done in the {\sc atmo} experiments.   The simulation was run for 500 days, and the results are compared to the unmodified UM/Venot and {\sc minichem} results at the same output time .

\begin{figure*}
	\includegraphics[]{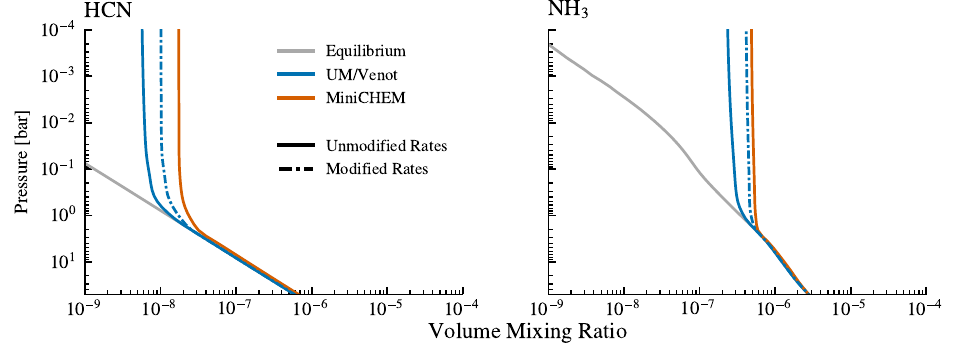}
    \caption{The area-weighted abundances for \ce{HCN} (left panel) and \ce{NH3} (right panel) for solar metallicity simulations, each run for 500 days. The solid lines are the {\sc minichem} (red) simulation and the UM/Venot (blue) simulation with full integration, both using their original reaction rates.  The dash-dot line is a UM/Venot simulation with two reaction rates changed, as discussed in the text.}
    \label{Fig:rate_test}
\end{figure*}

Figure \ref{Fig:rate_test} shows the area-weighted \ce{HCN} and \ce{NH3} abundances.  Compared to the previous UM/Venot simulation and as expected from the previous {\sc atmo} experiments, the simulation with the new rates now shows better agreement with the {\sc minichem} \ce{NH3} abundances, and the \ce{HCN} abundance also exhibits an increased abundance, moving it closer to agreement with {\sc minichem} although the change is not to the same degree as is seen in the \ce{NH3} abundances.   This supports our view that the use of the \citet{dean_1984} rate in {\sc vulcan} results in the differences in the quenched \ce{NH3} abundances, as well as explains part of the differences in the quenched \ce{HCN} abundances.

\section{The Chemical Timestep}
\label{Appendix:Timestep}

All simulations presented in the main text were done using a chemical timestep of $\Delta t_\mathrm{chem} = 3750\,\mathrm{s}$.  While this timestep is adopted in \citet{drummond_2020,zamyatina_2023,zamyatina_2024}, all of which use the UM chemical solver\footnote{Tests of the UM solver with a single reaction converting \ce{H} and \ce{H2} were performed in \citet{drummond_2017} and found a timestep of 6000~s could result in relative difference of 0.1\% in molar abundance relative to tests with chemical timesteps of 600~s and 60~s. This was one of the first tests of the UM solver during its development; however, this only investigates a single reaction and predates the use of the \citetalias{venot_2019} network with the UM.}, this choice is not universal in the literature, and the {\sc minichem} simulations presented in \citet{lee_2023}, for example, adopted a shorter timestep of $1500\,\mathrm{s}$.  To determine what impact this may have on the results presented here, we run tests of the $10\times$ solar metallicity case using the shorter chemical timestep of 1500~s, with the UM/Venot solver integrating for the full chemical timestep.   Each simulation is run for 300 days, sufficient to spin up the jet and exhibit quenching behaviour.  We focus on the $10\times$ solar metallicity case as this higher metallicity case will, in general, have the shortest chemical timescales and thus be the most likely to show effects of an overly large $\Delta t_\mathrm{chem}$.

Comparing first the horizontally-averaged abundances at day 300 for the UM/Venot with the differing $\Delta t_\mathrm{chem}$, the \ce{CH4}, \ce{H2O}, \ce{NH3}, \ce{CO}, \ce{HCN}, and \ce{CO2} abundances differ by less than 0.7\% between the simulations.  The abundances of \ce{H} and \ce{OH} radicals, which generally vary on shorter chemical timescales, differ by less than 3\%.   The {\sc minichem} simulations with differing chemical timesteps exhibit similar relative errors between the simulations. 

\begin{figure*}
	\includegraphics[]{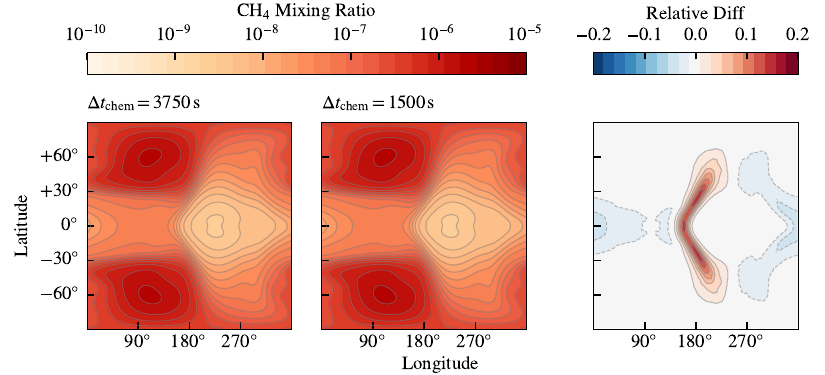}
    \caption{ {\em Left and middle panels: } The local \ce{CH4} volume mixing ratio at 0.1 bar for $10\times$ solar metallicity for two UM/Venot simulations using different chemical timesteps. {\em Right panel: } The relative difference between the $\Delta t_\mathrm{chem} = 3750\,\mathrm{s}$ and $\Delta t_\mathrm{chem} = 1500\,\mathrm{s}$ simulations, as computed as $\left(x_\mathrm{CH_4, 3750\,\mathrm{s}} - x_\mathrm{CH_4,1500\,\mathrm{s}}\right)/x_\mathrm{CH_4,1500\,\mathrm{s}}$, where $x_\mathrm{CH_4,\Delta t_\mathrm{chem}}$ is the \ce{CH4} volume mixing ratio using chemical timestep $\Delta t_\mathrm{chem}$. }
    \label{Fig:horiz_ch4}
\end{figure*}

If instead the simulations with differing chemical timesteps are compared at the level of individual cells within the computational domain, local errors are found to occur around sharp gradients in the chemical abundances.  These differences can be up to 40\%.   As an illustration, Figure \ref{Fig:horiz_ch4} shows the \ce{CH4} abundance for both UM/Venot simulations at 0.1 bar, as well as the relative difference between the two simulations.   Localised differences such as these around sharp gradients are unlikely to impact thermal and dynamic structures in the simulations, nor the synthetic observations, but it is nonetheless important to note that they exist. 

The impact of the shorter chemical timestep on the performance is significant, as for the UM/Venot simulation, the time to simulate one Earth day increases 891 s from 680 s, and for the {\sc minichem} simulation it increases to 616 s from 595 s.

\section{Opacity Feedback}
\label{Appdendix:Opacity}
The approach adopted here, as outlined in Section \ref{Sec:Chem}, is to have the chemical abundances inform the gas opacities, as this is both more physical and the how the chemistry schemes are used in the UM.  The temperatures differences that result from the choice of chemical scheme may feed back and exacerbate abundance differences between the schemes.  To investigate this possibility, we perform an additional test which runs both the UM/Venot and {\sc minichem} chemistry schemes within a single solar metallicity simulation (hereafter referred to as the ``two solver'' simulation).  The UM/Venot abundances are used to compute opacities while the {\sc minichem} abundances treated as passive tracers.  This implementation allows for the same gas temperatures to be used in each chemistry scheme.  While the UM/Venot abundances in this simulation will be the same as in the standard simulation, the {\sc minichem} abundances will differ from the standard {\sc minichem} simulation due to the temperature differences caused by the differing opacities. We ran this simulation for 500 days. 

\begin{figure}
	\includegraphics[]{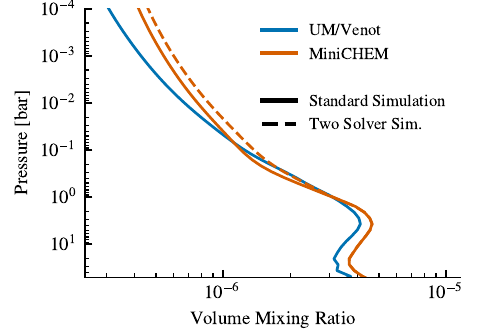}
    \caption{ The area-weighted \ce{CH4} volume mixing ratio the {\sc minichem} and UM/Venot simulations (solid lines), with the latter performing the full integration of the chemical timestep.  The dashed line shows the {\sc minichem} abundances from the ``two solver'' simulation in which the {\sc minichem} abundances are treated as passive tracers coupled to the UM/Venot simulation. Further details can be found in the text.  }
    \label{Fig:chem_two_solvers}
\end{figure}

The differences in the abundances between the standard {\sc minichem} simulation and two solver simulation are small, less than 2\%, often significantly less, except in the case of \ce{CH4}, where the area-weighted difference is up to 14\% at pressures less than 1 bar (see Figure \ref{Fig:chem_two_solvers}), with the abundances in the two solver simulation greater than in both the standard UM/Venot and {\sc minichem} simulations.  This highlights the sensitivity of the quenched \ce{CH4} abundance on the temperature structure of the atmosphere.

\section{Additional Plots}
\label{Appendix:Additional}

\begin{figure*}
	\includegraphics[]{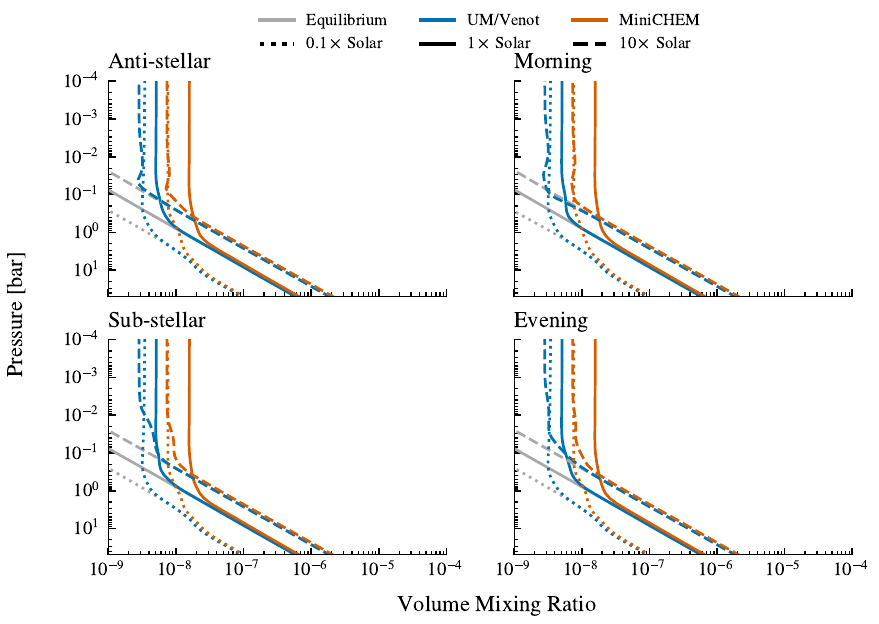}
    \caption{The volume mixing ratios for \ce{HCN} for equilibrium as well as the UM/Venot and {\sc minichem} chemical kinetics simulations. The UM/Venot simulations do not employ the escape condition. We omit the UM/Venot simulations using the escape condition to avoid figure overcrowding.}
    \label{Fig:chem_hcn}
\end{figure*}

\begin{figure*}
	\includegraphics[]{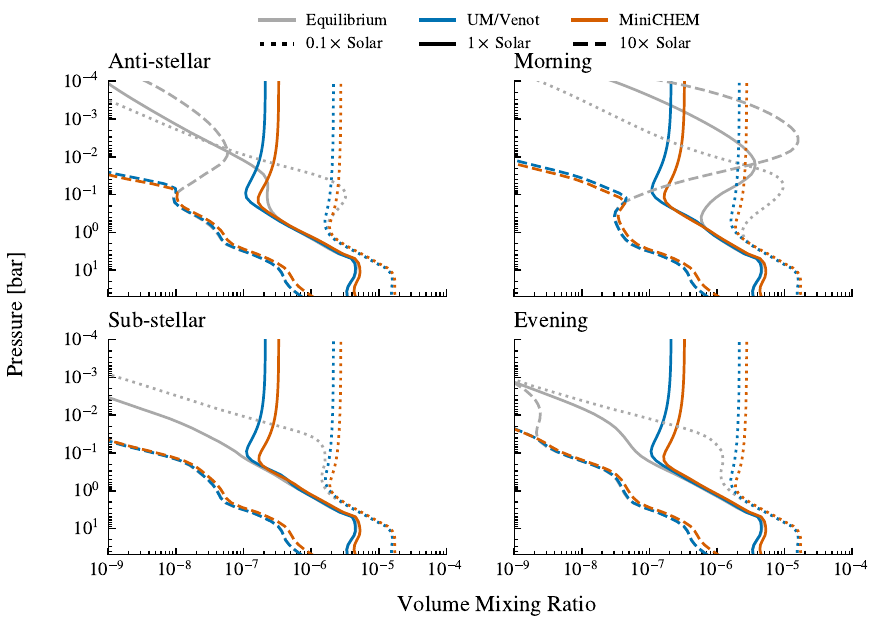}
    \caption{The volume mixing ratios for \ce{CH4} for equilibrium as well as the UM/Venot and {\sc minichem} chemical kinetics simulations. The UM/Venot simulations do not employ the escape condition. }
    \label{Fig:chem_ch4}
\end{figure*}

\begin{figure*}
	\includegraphics[]{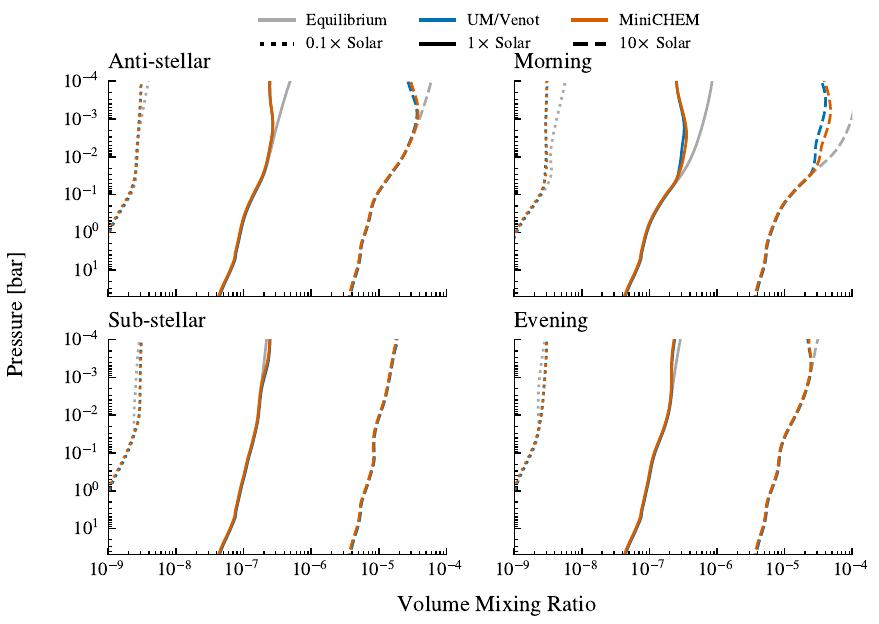}
    \caption{The volume mixing ratios for \ce{CO2} for equilibrium as well as the UM/Venot and {\sc minichem} chemical kinetics simulations. The UM/Venot simulations do not employ the escape condition. }
    \label{Fig:chem_co2}
\end{figure*}

\begin{figure*}
	\includegraphics[]{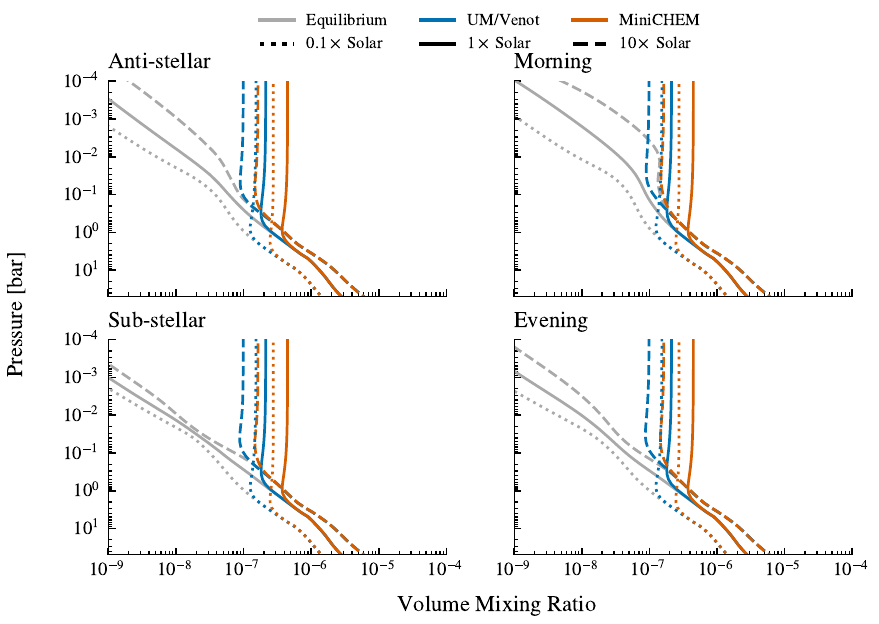}
    \caption{The volume mixing ratios for \ce{NH3} for equilibrium as well as the UM/Venot and {\sc minichem} chemical kinetics simulations. The UM/Venot simulations do not employ the escape condition. }
    \label{Fig:chem_nh3}
\end{figure*}

\begin{figure*}
	\includegraphics[]{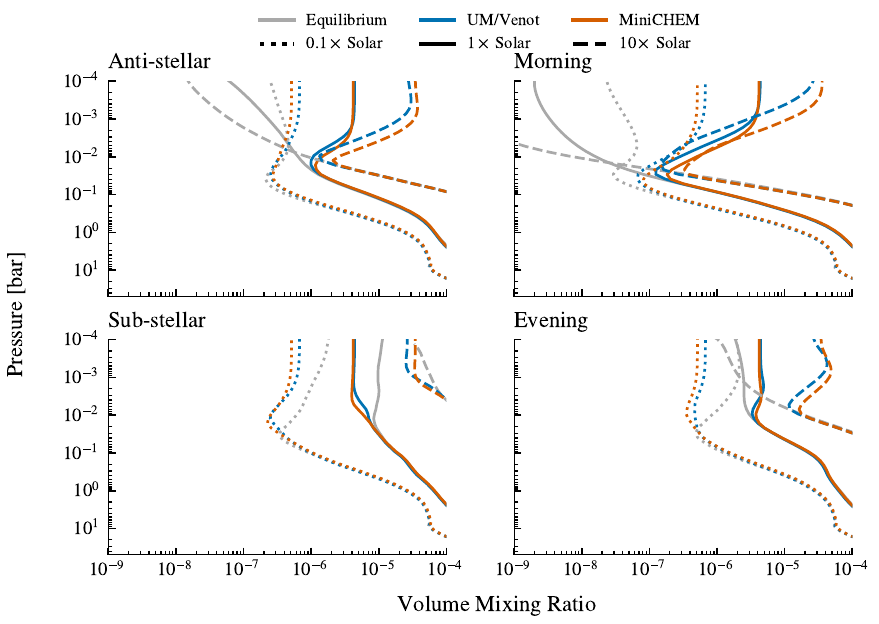}
    \caption{The volume mixing ratios for \ce{H} for equilibrium as well as the UM/Venot and {\sc minichem} chemical kinetics simulations. The UM/Venot simulations do not employ the escape condition. }
    \label{Fig:chem_h}
\end{figure*}

\begin{figure*}
	\includegraphics[]{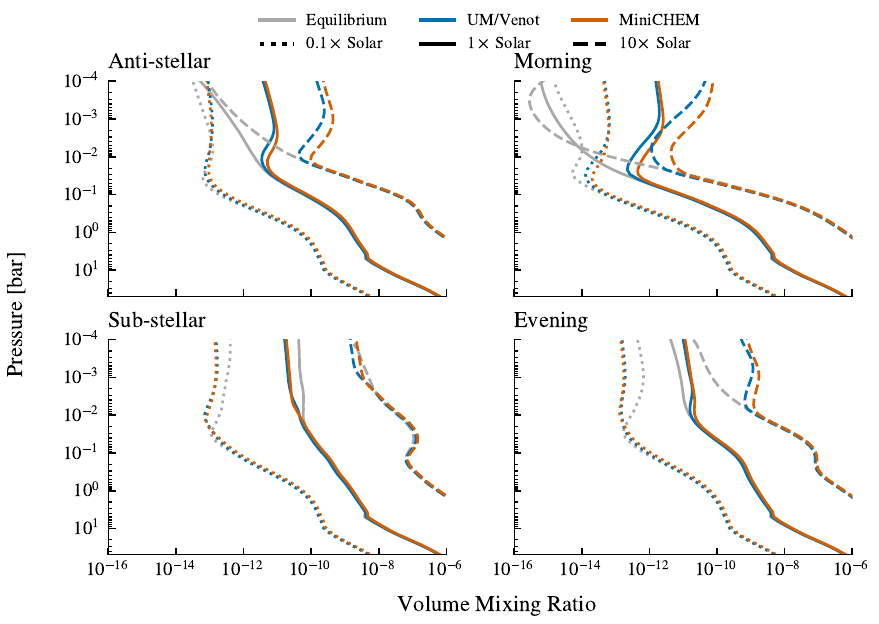}
    \caption{The volume mixing ratios for \ce{OH} for equilibrium as well as the UM/Venot and {\sc minichem} chemical kinetics simulations. The UM/Venot simulations do not employ the escape condition.  Note that due to the low abundances in the upper atmosphere, the range of values on the x axes is different from previous plots.}
    \label{Fig:chem_oh}
\end{figure*}

While in the main body of the text, we focused on differences in abundances averaged over isobars, we include here equatorial profiles along different directions for reference to highlight how abundance profiles vary around the equator.  The profiles for the four opacity sources focused on in the main text are in Figures \ref{Fig:chem_hcn} to \ref{Fig:chem_nh3} while the \ce{H} and \ce{OH} profiles are in Figures \ref{Fig:chem_h} and \ref{Fig:chem_oh}.

\bsp	
\label{lastpage}
\end{document}